\documentclass[reprint,superscriptaddress,aps,longbibliography]{revtex4-2}
\usepackage[utf8]{inputenc}
\usepackage[T1]{fontenc}
\usepackage{amsmath}
\usepackage{amssymb}
\usepackage{graphicx}
\usepackage[capitalize, nameinlink]{cleveref}
\usepackage{subfigure}
\graphicspath{{figures/}}
\usepackage{adjustbox}
\usepackage{booktabs}
\usepackage{multirow}

\usepackage{color}
\usepackage{rotating}
\usepackage{ulem}

\begin{document}

\title{Ferromagnetic resonance modes in trilayer artificial spin ices subject to interfacial Dzyaloshinskii-Moriya interaction}

\author{V. Vanga}
\affiliation{Center for Magnetism and Magnetic Nanostructures, University of Colorado Colorado Springs, Colorado Springs, CO, USA}
\author{G. Alatteili}
\affiliation{Center for Magnetism and Magnetic Nanostructures, University of Colorado Colorado Springs, Colorado Springs, CO, USA}
\author{E. Iacocca}
\email{eiacocca@uccs.edu}
\affiliation{Center for Magnetism and Magnetic Nanostructures, University of Colorado Colorado Springs, Colorado Springs, CO, USA}

\date{\today}

\begin{abstract}
Artificial spin ices are metamaterials that can host several ferromagnetic {resonances} as well as spin waves. As the field advances towards the creation of three-dimensional geometries, a trilayer square artificial spin ice has been already found to exhibit many interesting properties. Here, we numerically investigate a strongly-coupled trialyer square artificial spin ice under the effect of interfacial Dzyaloshinskii-Moriya interaction (DMI). This interaction affords non-reciprocity to waves, leading to changes in the standing wave modes established in confined geometries. We find that the interplay between the non-reciprocity, an applied field, and the stray field within the artificial spin ice results in {frequency} split additional edge modes. The edge modes are favored by the DMI sign and exhibit destructive and constructive interference depending on both the DMI magnitude {and the external magnetic field}. Our results demonstrate the non-reciprocity in small nanoislands can affect the long-range states stabilized in the {artificial spin ice} due to {the} strong coupling between layers.
\end{abstract}
\maketitle

\section{Introduction}

The control and manipulations of magnons, the quanta of angular momentum in magnetic materials, is central to the field of magnonics~\cite{Krawczyk2014,Chumak2022}. A common method to realize magnonic crystals is to pattern a super-lattice leading to the development of a magnon band structure due to Bloch's theorem~\cite{Nikitov2001,Neusser2009,Wang2010,Kumar2014,Gubbiotti2018,Lisiecki2019,Szulc2022,Negrello2022,Roxburgh2024,Roxburgh2025}. In two-dimensions, magnonic crystals take the form of nanodots~\cite{Tacchi2011} and antidot~\cite{Neusser2011} lattices. A natural extension is to consider anisitropic structures as the composing elements, whose arrangements are known as artificial spin ices~\cite{Gliga2020}. While artificial anti-spin-ices have shown clear bandgaps~\cite{Mamica2018}, band structure in artificial spin ices (ASIs) remains challenging despite theoretical predictions~\cite{Iacocca2016,Wysin2013,Lasnier2020} due to limited coupling between elements. ASIs are reconfigurable~\cite{Heyderman2013} and their use as magnonic crystals would additionally bring about the benefit of a reconfigurable band structure~\cite{Gliga2020}.

In ASIs, the geometric arrangement of nanomagnets dictates their static properties, such as magnetic states~\cite{Heyderman2013,Skjaervo2020}. When considering GHz dynamics~\cite{Lendinez2019,Gliga2020}, it is essential to distinguish between the ASI's static and dynamic coupling. On the one hand, the static coupling is given by the stray fields of the nanomagnets that act as an effective bias field on other nanomagnets. This gives rise to magnetization-state-dependent ferromagnetic resonance (FMR)~\cite{Gliga2013,Jungfleisch2016,Arroo2019,Dion2019,Vanstone2022,Lendinez2023,Alatteili2024} which can be routinely measured electrically, optically, or both~\cite{Lendinez2019}. The sensitivity of FMR to the underlying magnetization state has been exploited for reservoir computing quite successfully~\cite{Gartside2022,Lee2024,Stenning2024}. On the other hand, the dynamic coupling is due to the magnetization dynamics and are thus fluctuations in the stray field. The spectral features of those dynamics allow modes to couple and{, therefore,} the dynamic coupling can give rise to a magnonic band structure. Unfortunately, the dynamic coupling is extremely weak, limiting the spectral features of the band structure~\cite{Alatteili2023,Alatteili2025}. Indeed, it has been demonstrated in macroscopic ASIs that band structure~\cite{Scafuri2025} and frequency combs~\cite{Peroor2025} arise due to the coupling between the stray fields of bar magnets and their dynamic mechanical motion. Therefore, it is critical to increase the dynamic coupling in ASIs to realize reconfigurable band structures and nonlinear phenomena.

There have been several ASI geometries that can result in an increased dynamic coupling. One of the first examples is the vertex-modified ASI~\cite{Ostman2018} in which nanoparticles are placed at the vertices to mediate coupling. Another example {is} to stack nanomagnets to enhance interactions by vertical proximity~\cite{Berchialla2024,Sultana2025,Szulc2022,Micaletti2025,Micaletti2025b}. Such a trilayer ASI~\cite{Dion2024} already demonstrated ultrastrong magnon-magnon coupling in FMR. In that work, modes with parallel and antiparallel geometry were selected based on the field orientation and a slight structural asymmetry in the trilayer.
\begin{figure}[b]
\includegraphics[trim={0 0.6in 0 0.5in}, clip, width=3in]{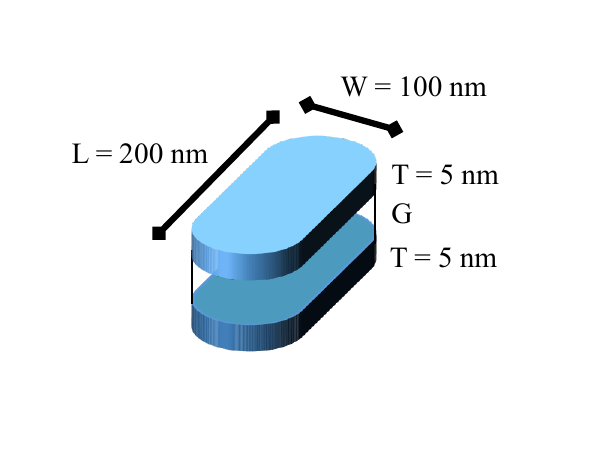}
\caption{Schematic of a trilayer element composed of a top Py layer (light blue) and a bottom CoFe layer (dark blue) separated by a nonmagnetic spacer of thickness $G$. The stadium-shaped nanoislands have dimensions $L = 200$~nm, $W = 100$~nm, and $T = 5$~nm.}
\label{fig:geometry}
\end{figure} 

Another way to induce asymmetry in spin waves is through the Dzyaloshinskii-Moriya interaction (DMI)~\cite{Nembach2015}. The nonreciprocity originating from DMI is known to give rise, e.g., to caustic waves~\cite{Kim2016c} and band flattening in magnonic band structure~\cite{Gallardo2019}. The effect of nonreciprocity has been also investigated in confined geometries~\cite{Zingsem2019}. Here, we investigate the effect of DMI in a trilayer ASI by micromagnetic simulations. It is found that additional edge modes arise due to the interplay between {DMI} and magnetic fields within the ASI. In particular, transverse or longitudinal edge modes are favored by the DMI sign while constructive and destructive interference is encouraged by the combination of DMI magnitude and externally applied field. These results suggest further investigation in 3D spin ices where at least one layer is interfaced with a heavy metal as well as the possibility to afford topological protection to spin waves.
\begin{figure}[t]
\includegraphics[trim={0 0.3in 0 0.5in}, clip, width=3.3in]{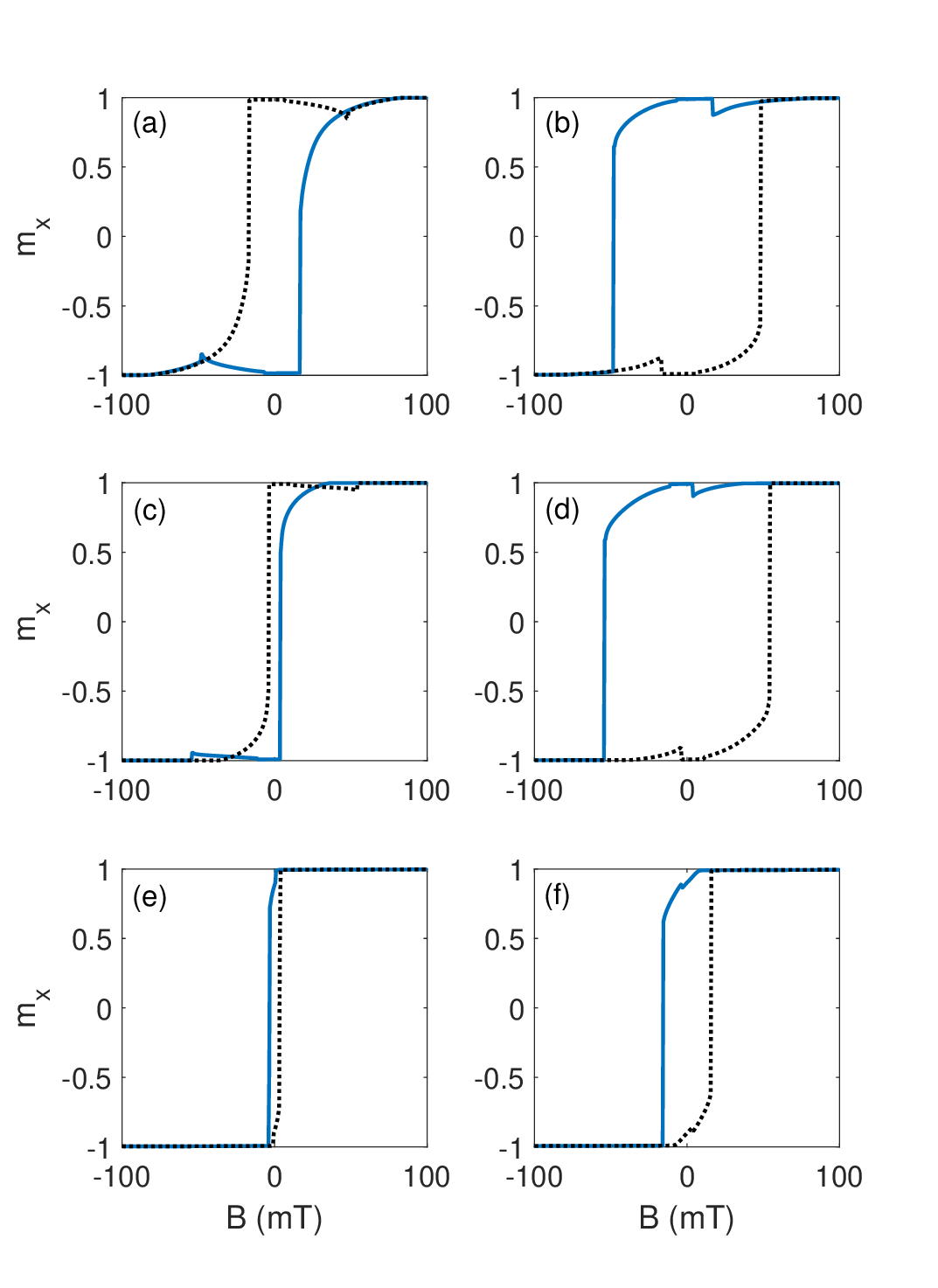}
\caption{Hysteresis loops for a trilayer element for gaps of (a-b), $5$~nm, (c-d) $20$~nm, and (e-f) $40$~nm. The hysteresis curves of Py are shown in (a), (c), and (e) while the hysteresis curve of CoFe are shown in (b), (d), and (f). In all cases, the field sweep from positive to negative field is shown by blue curves and the sweep from negative to positive field is shown by black dashed curves. }
\label{fig:Hystersis}
\end{figure} 

\section{Single trilayers}
\label{sec:single}

We first characterize the magnetostatics and dynamics of a single trilayer element. In order to distinguish the features of each layer, we consider a trilayer composed of a top permalloy (Py) layer and a bottom CoFe layer. The material parameters for Py are: saturation magnetization $M_s = 790$ kA/m, exchange constant $A = 10$ pJ/m, and Gilbert damping constant $\alpha = 0.01$; and for CoFe are~\cite{Schoen2016}: saturation magnetization $M_s = 1700$ kA/m, exchange constant $A = 26$ pJ/m, and Gilbert damping constant $\alpha = 0.0004$.

The trilayers are simulated by the micromagnetic software MuMax3~\cite{Vansteenkiste2014}. The nanomagnets are considered to have a stadium-shaped cross-section composed of a rectangular bulk section capped with semicircular sections on each end~\cite{Strandqvist2025,Vanga2025}. The trilayer geometry is depicted in Fig.~\ref{fig:geometry}. The nanoislands' length is $L = 200$~nm, width $W = 100$~nm, and thickness $T = 5$~nm. The system is discretized to $128\times 64 \times G/T$, cells where the gap $G$ is discretized in factors of $T$. This leads to a cell size of 1.56~nm~$\times$~1.56~nm~$\times~T$.
\begin{figure}[t]
\includegraphics[trim={0 0.3in 0 0.5in}, clip, width=3.3in]{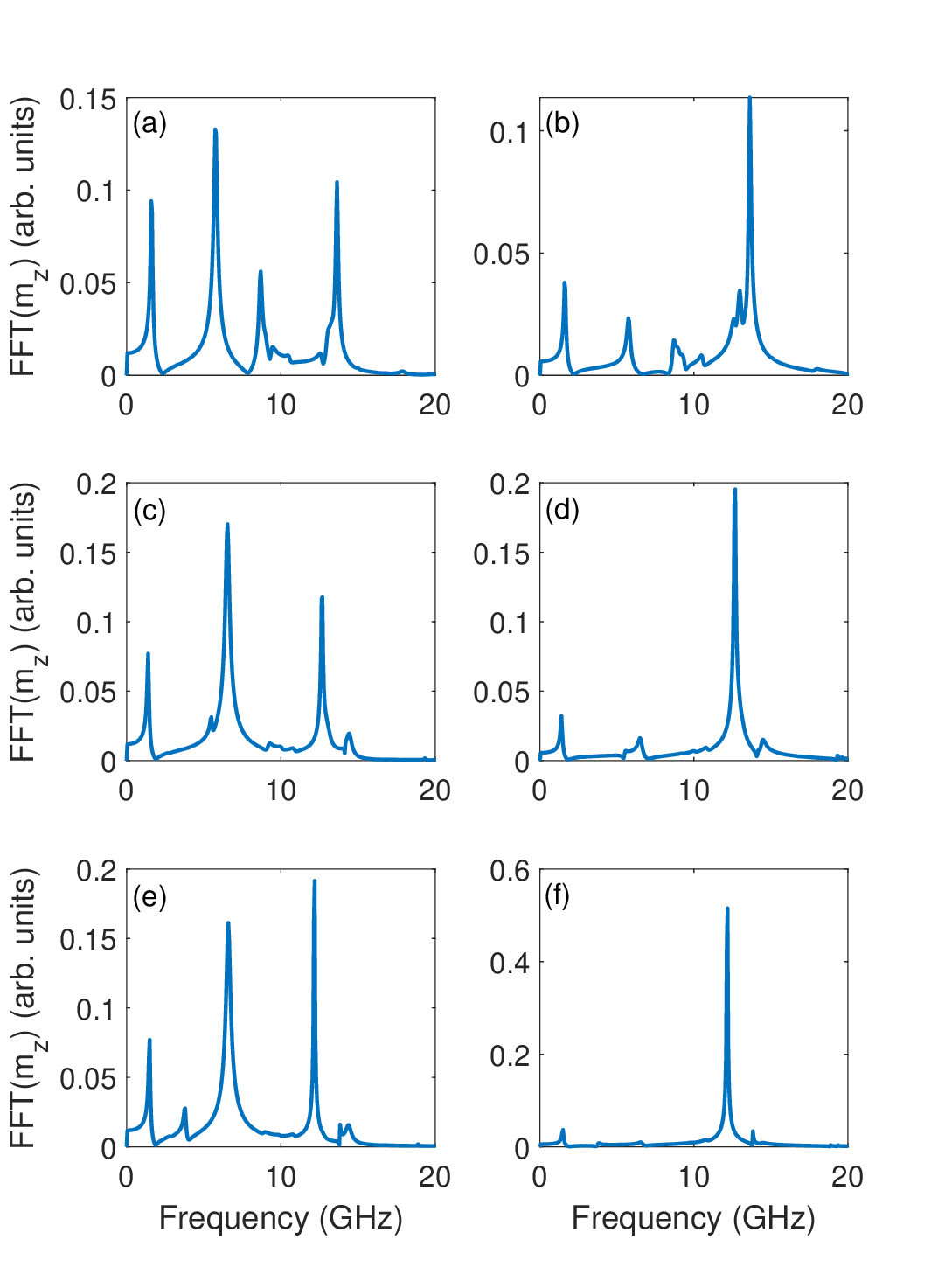}
\caption{Ferromagnetic resonance for a trilayer element for gaps of (a-b), $5$~nm, (c-d) $20$~nm, and (e-f) $40$~nm. The FMR of Py are shown in (a), (c), and (e) while the FMR of CoFe are shown in (b), (d), and (f).}
\label{fig:SingleTrilayerSpectra}
\end{figure}

To assess the coupling strength between the layers, we compute the hysteresis loops at different gap distances $G$. The {external field $B$}, is swept in the range [-100~mT, 100~mT] using a 0.5~mT step along the trilayer's easy axis. The hysteresis loop for each material is shown in Fig.~\ref{fig:Hystersis} as a function of $G$. We use gaps of (a-b) $G=5$~nm, (c-d) $G=20$~nm, and (e-f) $G=40$~nm. We show the hysteresis loop for the Py layer in panels (a), (c), and (e), and the CoFe layer in panels (b), (d), and (f). In all cases, the blue curve indicates the field sweeping from positive to negative and the black dashed curves indicates the field sweeping from negative to positive. When $G=5$~nm, Fig.~\ref{fig:Hystersis}(a) and (b), the system is strongly coupled, as evidenced by the many features in the hysteresis loop that deviate from the standard Stoner-Wohlfarth model. In particular, we see that the Py has no coercivity because of the CoFe stray field that opposes the external field and leads to a premature reversal. At that moment, the stray field from the Py partly compensates the external field, leading to the CoFe experiencing a smaller field magnitude and an {increase} in its net magnetization. As the external field magnitude {continues to decrease and become negative}, the CoFe continues its reversal process while the effective field in the Py is reduced in magnitude. This causes the magnetization increase in the Py. Finally, when the CoFe switches, the trilayer is in a parallel state along the external field and saturation is eventually achieved. Similar features are observed for $G=20$~nm in Fig.~\ref{fig:Hystersis}(c) and (d), however their magnitude is reduced. Finally, for $G=40$~nm shown in Fig.\ref{fig:Hystersis}(e) and (f), the hysteresis loops are almost ideal, indicating the the system can be considered to be decoupled.
\begin{figure}[b]
\includegraphics[trim={0 0in 0 0in}, clip, width=3in]{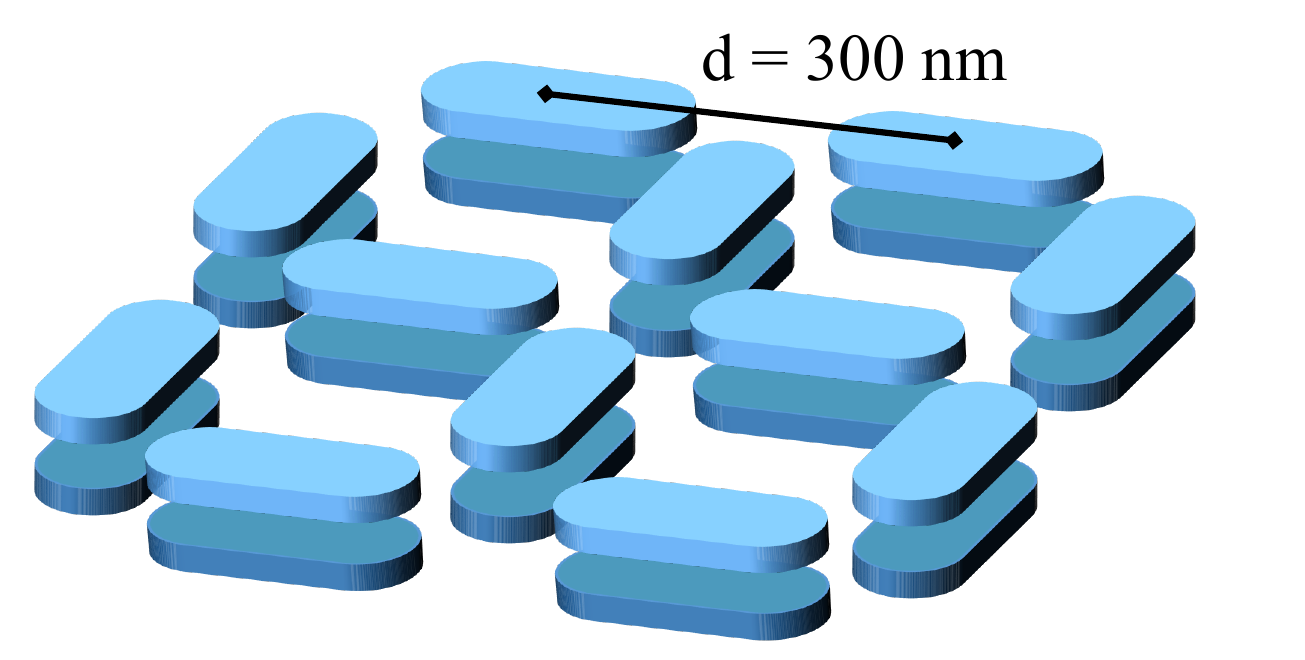}
\caption{Schematic of a trilayer square ASI where each element is the trilayer shown in Fig.~\ref{fig:geometry}. The center-to-center distance is $d=300$~nm.}
\label{fig:ASIgeometry}
\end{figure} 

The dynamic coupling between the layers can be evidenced through their ferromagentic resonance (FMR) response. We set our initial state at $B=0$~T and an antiparallel orientation between nanomagnets, which is left to relax for $10$~ns to eliminate spin wave resonances. To numerically compute the FMR, we applied a 100~mT field pulse normal to the element's plane for 10~ps and recorded the response of the system for $40$~ns sampled at 10~ps. These parameters give a frequency resolution of 25~MHz and a maximum frequency of {50}~GHz. The results are shown in Fig.~\ref{fig:SingleTrilayerSpectra} as a function of $G$, c.f.~Fig.~\ref{fig:Hystersis}. We use gaps of (a-b) $G=5$~nm, (c-d) $G=20$~nm, and (e-f) $G=40$~nm where the FMR for Py is shown in panels (a), (c), and (e), and that for CoFe layer in panels (b), (d), and (f). The clearest evidence of decoupling{, as G increases,} is seen by the CoFe layer. We see several peaks for $G=5$~nm in panel (b) and only one prominent peak for $G=40$~nm in panel (f). The difference in damping constants also results in different linewidths. This allows us to distinguish between modes on each layer. For example, in panel (e), we recognize the 12.18~GHz narrow peak that is driven by the CoFe layer compared to the Py own peak at 6.6~GHz. Because the saturation magnetization of Py is lower than CoFe, the opposite is not true. In other words, the $6.6$~GHz peak in panel (f) is not clearly seen. For small gap distances, an increase in coupling is evidenced by the {large number of} peaks, frequency shifts, and the comparable linewidths. For $G=5$~nm, panels (a) and (b), it is difficult to discern the origin of the peaks.

From the combination of hysteresis and FMR calculations, we verified that $G=5$~nm ensures a strong static and dynamic coupling in the trilayer. Therefore, we will use only this spacing for the ASI.
\begin{figure}[t]
\includegraphics[trim={0 0.5in 0 0.2in},clip, width=3.5in]{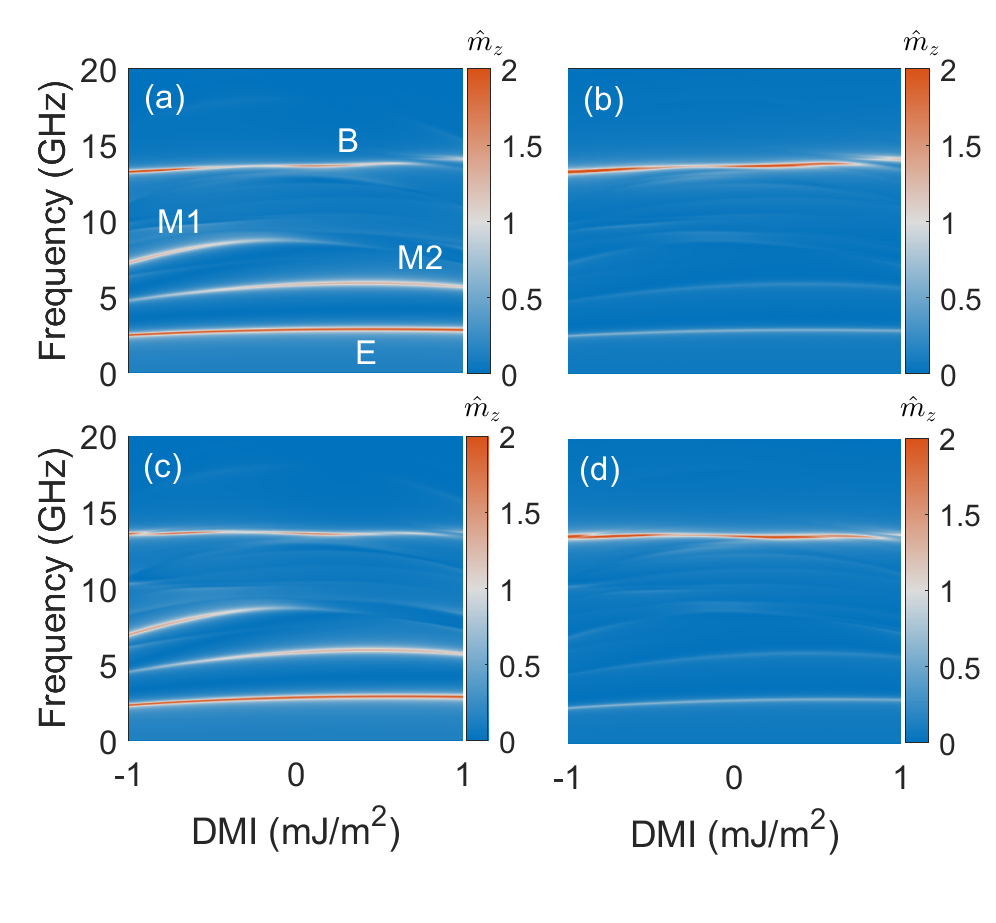}
\caption{DMI-dependent spectra for trilayer ASIs for the cases in which DMI has the same sign for (a) Py and (b) CoFe; and DMI is opposite values for (c) Py and (d) CoFe. In the latter pannels, the sign of the x-axis represents the DMI parameters for the Py layer. For dominant modes are identified: bulk (B), edge (E), and two standing-wave modes, M1 and M2.}
\label{fig:ASI_DMI}
\end{figure} 
\begin{figure*}[t]
\includegraphics[trim={0 0.6in 0 0.3in},clip,width=\linewidth]{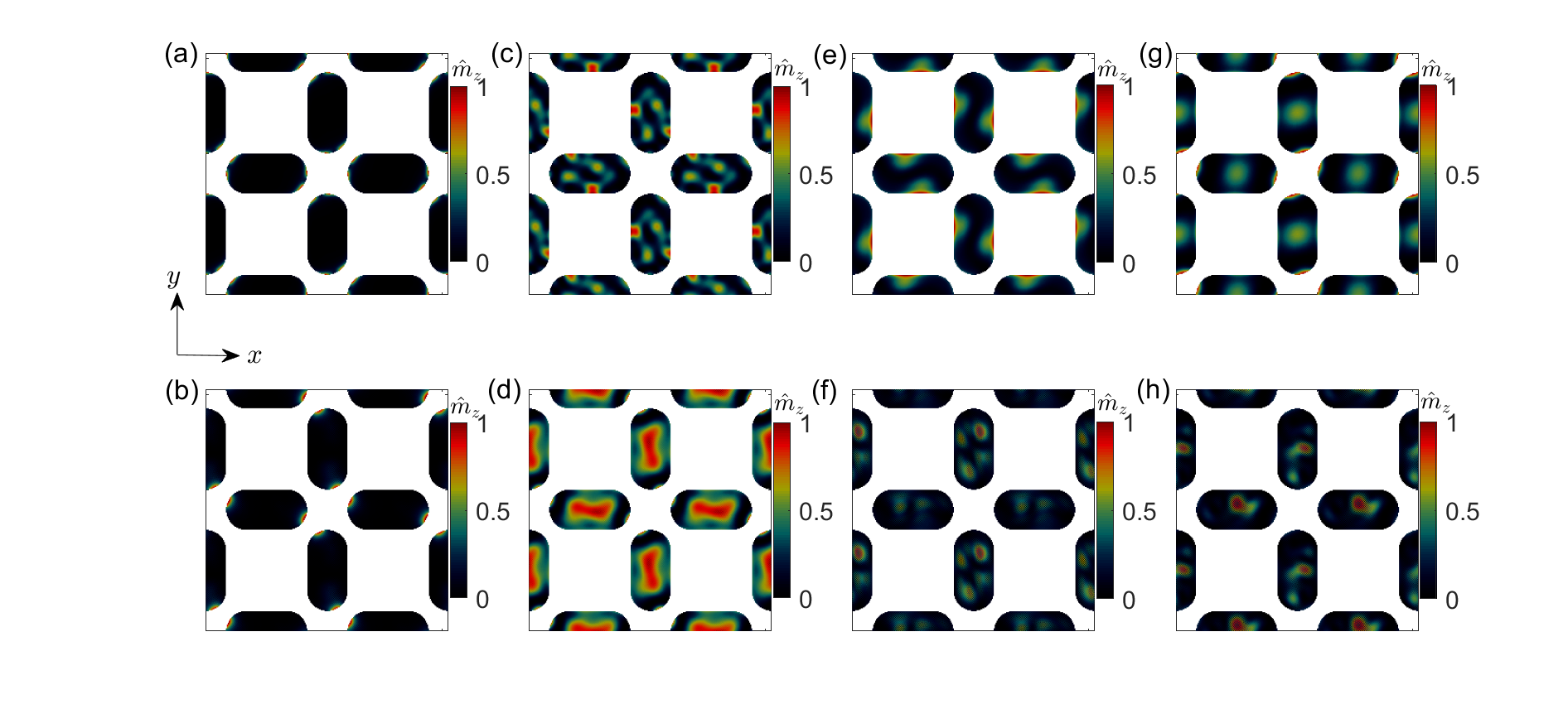}
\caption{Mode profiles as a function of DMI. The modes for Py are shown in panels (a), (c), (e), and (g) while modes for CoFe are shown in panels (b), (d), (f), and (h). The edge mode E is shown in (a) and (b) for $D=-1$~mJ/m$^2$. The bulk mode B is shown in (c) and (d) for $D=-1$~mJ/m$^2$. The M1 mode is shown in (e) and (f) for $D=-1$~mJ/m$^2$. The M2 mode is shown in (g) and (h) for $D=1$~mJ/m$^2$. In all cases, the modes' amplitude has been normalized. The reference frame is indicated in the figure. Each panel shows a $600$~nm$\times600$~nm area.}
\label{fig:ASI_Modes_DMI}
\end{figure*} 

\subsection{Trilayer Artificial Spin Ice}

We simulate a trilayer square ASI as depicted in Fig.~\ref{fig:ASIgeometry}. We use a center-to-center distance of $d=300$~nm and discretize the simulation into a 256~$\times$~256~$\times$~3 domain which achieves a cell size of 2.34~nm~$\times$~3.34nm~$\times$~5nm for a full physical domain of 600~nm~$\times$~600~nm~$\times$~15~nm. Periodic boundary conditions are set in the plane of the ASI, setting 100 repetitions along each dimension. The normal component is left as free spin. The cell size and periodic boundary conditions are sufficient to achieve a good frequency resolution, as demonstrated earlier~\cite{Iacocca2020,Vanga2025}. DMI is introduced in both layers and we distinguish between cases where the DMI sign is the same or opposite {between the Py and CoFe layers} as DMI+ and DMI-, respectively.

The FMR of the ASI is computed using the same method and parameters described in section~\ref{sec:single}. We initialize the spin ice in an antiparallel remanent state, which can be accessed by an external magnetic field. The DMI is varied in from $D=0$ to $1$~mJ/m$^2$ in steps of 0.01~mJ/m$^2$. For each DMI value, the simulation is allowed to relax and then the FMR is computed. The DMI is swept to positive and negative values from zero to ensure a smooth change in the initial magnetization state. The spectra are then composed in a single plot with a variation from negative to positive DMI values.

The DMI-dependent spectra for the DMI+ case is shown in Fig.~\ref{fig:ASI_DMI}(a) for Py and (b) for CoFe. The predominant features are the two modes in between 5~GHz and 10~GHz exhibiting a strong DMI-dependence. We identify them as M1 and M2 in the figures. M1 is dominant for negative DMI while M2 is dominant for positive DMI. Furthermore M1 exhibits a strong red-shift with DMI while M2 varies only slightly. The bulk and edge modes, B and E, exhibit a modest dependence to DMI. This is a clear indication that M1 and M2 are standing waves in the nanoislands, as seen in Ref.~\cite{Zingsem2019}, while B and E are the zeroth-order modes that appear due to space quantization. However, we see that B splits at DMI of about 1~mJ/m$^2$, suggesting a change in the magnetization state. A very similar picture is observed for the DMI- case, in which the x axis represents the DMI value in the Py, panel (c), and it is understood that the CoFe has an opposite DMI value, panel (d). Most modes exhibit the same tunability to DMI, except the bulk mode which appears flatter when the signs are different. This is indicative of variations in the equilibrium magnetization state in both cases.

To further understand the nature of these modes, we plot their mode profiles. For this, we perform dedicated simulations with the same parameters described before but with a time step of 20~ps, resulting in a Nyquist frequency of $25$~GHz. The modes are presented in Fig.~\ref{fig:ASI_Modes_DMI}: for $D=-1$~mJ/m$^2$, we show the edge mode {, E,} for (a) Py and (b) CoFe as well as the bulk mode, B, for (c) Py and (d) CoFe. These are familiar modes in square ASI. The edge mode is strongly localized at the edges and hardly visible. However, we distinguish the typical asymmetry associated with a remanent state. The bulk mode is primarily seen in the CoFe layer, panel (d), and is {more} extended than {that of typical square} ASIs. Notably, the bulk mode in the Py layer, panel (c), is {not well defined, and appears to be localized at a lateral edge}. This is indicative that DMI has a stronger influence in the static magnetization state for Py due to the combination of a low $M_s$ and the stray field from CoFe.

The M1 mode is shown in (e) for Py and (f) for CoFe. We see that the mode exists primarily in the Py layer and corresponds to a transverse standing wave. In contrast, the M2 mode shown in (g) for Py and (h) for CoFe is a combination of ``typical'' edge and bulk modes that can be thought of as a longitudinal standing wave. In both cases, the modes are virtually inexistent in the CoFe layer, also seen from Figs.~\ref{fig:ASI_DMI}(b) and (d). The symmetries of these modes are indicative of the interplay between stray fields and DMI. The Py exhibits a much stronger ``buckling'' of the magnetization at the edges of the nanoislands. This enables modes to more strongly oscillate and establish standing waves. The CoFe has a more uniform magnetization states, in which the typical edge and bulk modes can be excited.
\begin{figure}[t]
\includegraphics[trim={0 0.4in 0 0.2in},clip, width=3.5in]{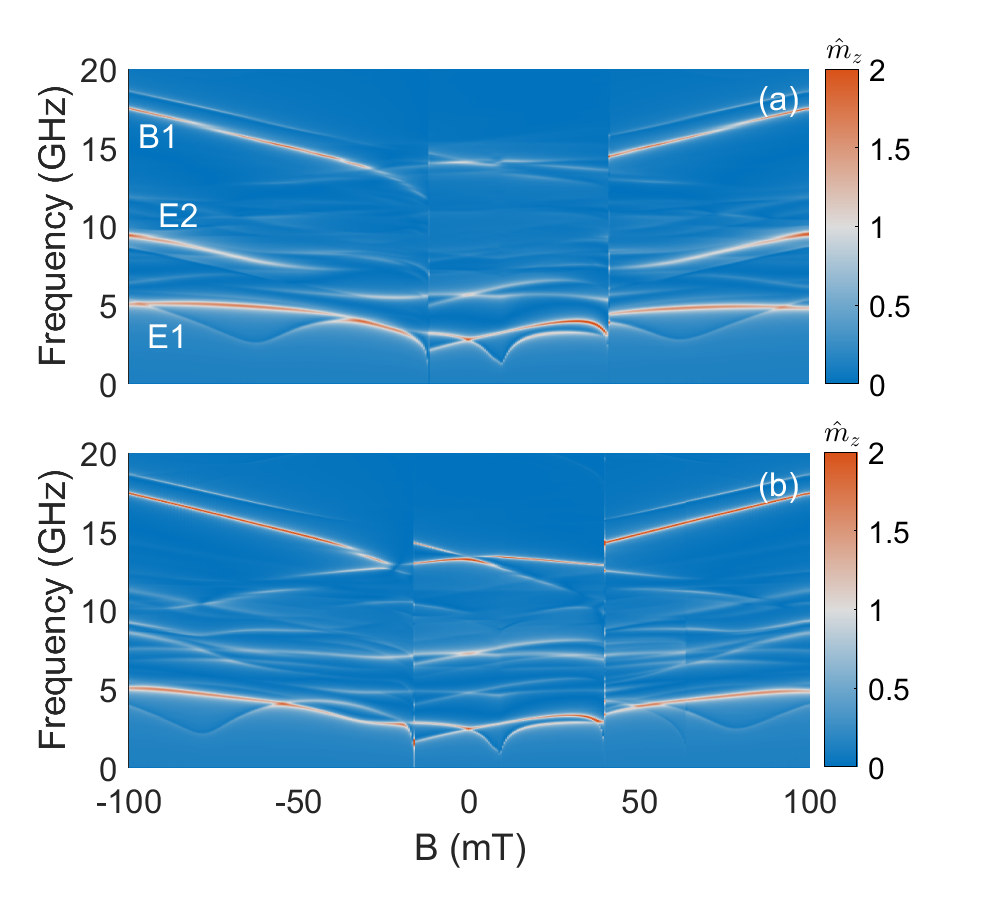}
\caption{Field-dependent spectra for trilayer ASIs when the DMI has the same sign for both trilayers and we choose (a) $D=1$~mJ/m$^2$ and (b) $D=-1$~mJ/m$^2$. The spectra is obtained from the average magnetization on the whole trilayer. In (a), modes E1, E2, and B1 are identified at saturation. The bulk mode at remanence has a weak amplitude due to destructive interference. In (b), E2 is split into several weaker modes while the bulk mode at remanence is present.}
\label{fig:ASI_Field}
\end{figure} 

We now turn to the spectra as a function of field in Fig.~\ref{fig:ASI_Field} for the cases of (a) $D=1$~mJ/m$^2$ and (b) $D=-1$~mJ/m$^2$. We choose only the DMI+ configuration in which the DMI sign is the same for both the Py and CoFe based on the previous results. In Fig.~\ref{fig:ASI_Field}, the spectra is obtained from the average magnetization in the full trilayer and the field is swept from negative for positive. At the condition $B=0$, we see the same modes observed in Fig.~\ref{fig:ASI_DMI}. These modes split for any non-zero field. The field dependence at remanence is reminiscent of previous results in square ASIs~\cite{Lendinez2021,Lendinez2023}, e.g., strong redshifts that soften and recover due to the compensation between the ASI and external field. However, the main features of the trilayer is that modes split with different tunabilities because the external field competes with the ASI remanent field and the field in each trilayer.
\begin{figure}[t]
\includegraphics[trim={1.4in 1.5in 1in 1.2in},clip, width=1.9in]{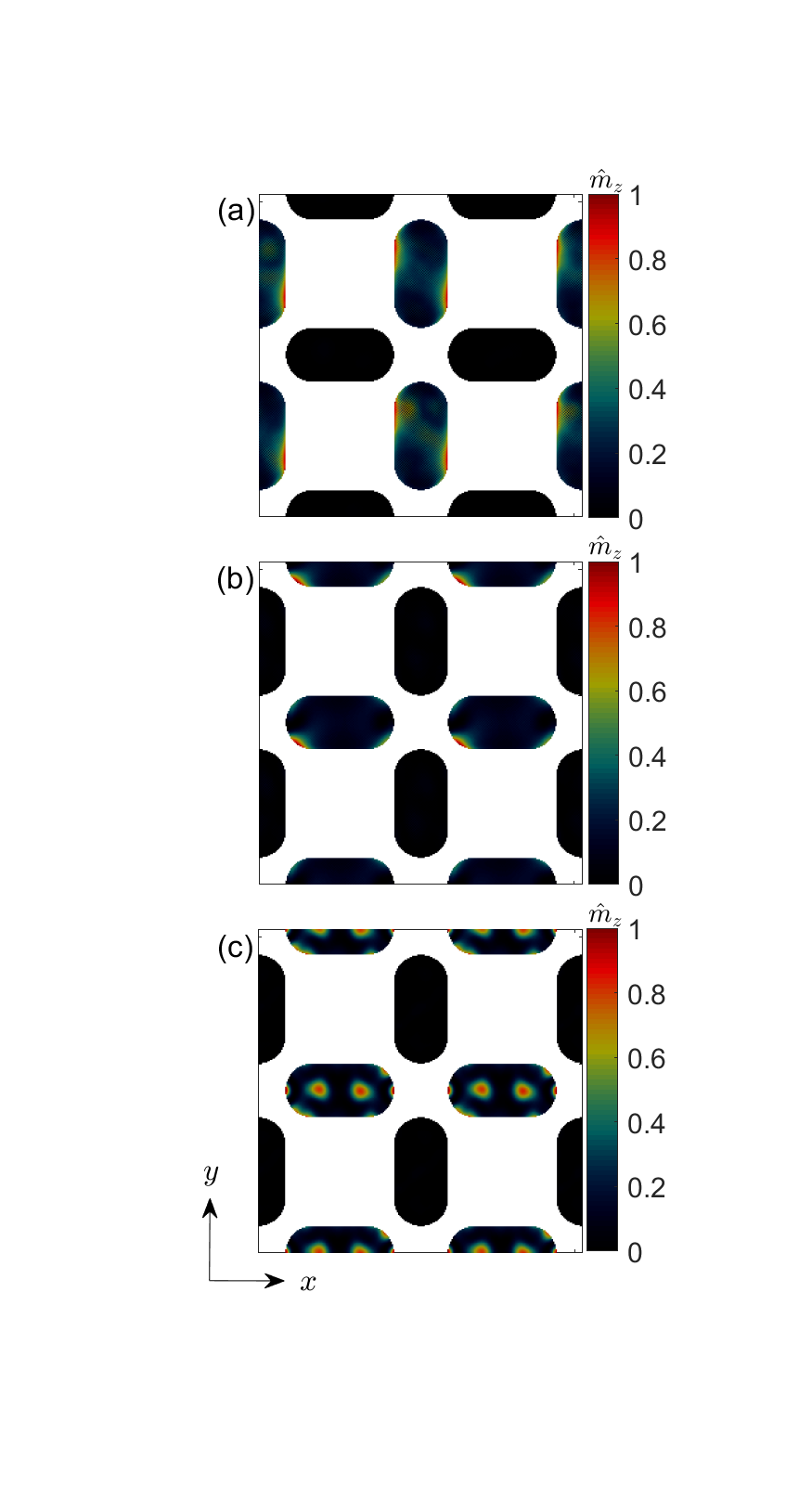}
\caption{Mode profiles for the saturating field $B=-100$~mT and $D=1$~mJ/m$^2$ for both layers. These mode profiles correspond to the identified modes in Fig.~\ref{fig:ASI_Field}, namely, (a) E1, (b) E2, and (c) B1. The reference frame is indicated in the figure. Each panel shows a $600$~nm$\times600$~nm area.}
\label{fig:ASI_Modes_Field}
\end{figure} 

Several features can be identified. It is now apparent that the bulk mode for the $D=1$~mJ/m$^2$ case is weak and further suggests destructive interference, while the same mode for $D=-1$~mJ/m$^2$ has a large amplitude. It is also clear that the spectrum is much more split in the $D=-1$~mJ/m$^2$ case for all fields in between the dominant edge and bulk modes. Even when the ASI is magnetized at $B=\pm100$~mT, there are effectively two dominant peaks: and edge mode (E1) at $\approx5$~GHz and a bulk mode (B1) at $\approx17.5$~GHz. In contrast, there is a well-defined mode at $\approx10$~GHz when the ASIs are saturated {in the $D=1$~mJ/m$^2$ case}. It is assumed that this mode arises due to constructive interference that is otherwise absent in the $D=-1$~mJ/m$^2$ case. The reverse situation is seen for the bulk mode at remanence, with a vanishing amplitude in (a) and notable amplitude in (b). These features {are a result} of the mode selection imposed by non-reciprocity.

It is interesting to look at the mode profiles when the ASI is magnetized but subject to DMI. The modes E1, E2, and B1 are shown in Fig.~\ref{fig:ASI_Modes_Field}(a), (b), and (c), respectively. E1 is the expected edge mode when the magnetization is saturated along the $x$ axis. E2 is an edge mode excited along the saturated islands which has a higher frequency because of the larger internal field. Finally, the B1 mode is actually a standing wave, similar to the standing wave mode seen in Fig.~\ref{fig:ASI_Modes_DMI}(c).

\section{Conclusions}

We have investigated the FMR of a trilayer square ice composed of different materials on each layer and subject to DMI. A spacing of $5$~nm was chosen to enforce strong static and dynamic coupling between the nanomagnets. The DMI changes the static magnetization state, which translates into frequency shifts for the FMR modes, as is expected for nanomagnets~\cite{Zingsem2019}. In general, the field dependent magnetization looks similar {to} that observed for other square ASIs~\cite{Lendinez2019,Lendinez2023}. However, closer inspection shows that additional edge modes are enabled by DMI and occur in the softer Py layer. In addition, DMI also induces preferential phases that can lead to modes destructively interfering.

This work opens more possibilities to investigate FMR or spin waves dynamics in ASIs. Previous results have demonstrated that DMI can be used to produce tailored magnetization states~\cite{Luo2019} and numerically indicated the potential for topologically protected magnons~\cite{Iacocca2017c}. Entering into the 3D realm~\cite{Berchialla2024,Sultana2025}, the role of DMI can be relevant for further tuning of FMR responses and band structures potentially exhibiting topological protection. In our study, the CoFe layer resulted largely insensitive to DMI, suggesting that only the soft layer can be interfaced with a heavy metal layer to measure the predicted FMR behavior. 

\section*{Acknowledgments}

This material is based upon work supported by the National Science Foundation under Grant No. 2532497.

\section*{Data availability}

The data that support the findings of this article are openly available~\cite{OSF}.


\begin{thebibliography}{55}%
\makeatletter
\providecommand \@ifxundefined [1]{%
 \@ifx{#1\undefined}
}%
\providecommand \@ifnum [1]{%
 \ifnum #1\expandafter \@firstoftwo
 \else \expandafter \@secondoftwo
 \fi
}%
\providecommand \@ifx [1]{%
 \ifx #1\expandafter \@firstoftwo
 \else \expandafter \@secondoftwo
 \fi
}%
\providecommand \natexlab [1]{#1}%
\providecommand \enquote  [1]{``#1''}%
\providecommand \bibnamefont  [1]{#1}%
\providecommand \bibfnamefont [1]{#1}%
\providecommand \citenamefont [1]{#1}%
\providecommand \href@noop [0]{\@secondoftwo}%
\providecommand \href [0]{\begingroup \@sanitize@url \@href}%
\providecommand \@href[1]{\@@startlink{#1}\@@href}%
\providecommand \@@href[1]{\endgroup#1\@@endlink}%
\providecommand \@sanitize@url [0]{\catcode `\\12\catcode `\$12\catcode
  `\&12\catcode `\#12\catcode `\^12\catcode `\_12\catcode `\%12\relax}%
\providecommand \@@startlink[1]{}%
\providecommand \@@endlink[0]{}%
\providecommand \url  [0]{\begingroup\@sanitize@url \@url }%
\providecommand \@url [1]{\endgroup\@href {#1}{\urlprefix }}%
\providecommand \urlprefix  [0]{URL }%
\providecommand \Eprint [0]{\href }%
\providecommand \doibase [0]{https://doi.org/}%
\providecommand \selectlanguage [0]{\@gobble}%
\providecommand \bibinfo  [0]{\@secondoftwo}%
\providecommand \bibfield  [0]{\@secondoftwo}%
\providecommand \translation [1]{[#1]}%
\providecommand \BibitemOpen [0]{}%
\providecommand \bibitemStop [0]{}%
\providecommand \bibitemNoStop [0]{.\EOS\space}%
\providecommand \EOS [0]{\spacefactor3000\relax}%
\providecommand \BibitemShut  [1]{\csname bibitem#1\endcsname}%
\let\auto@bib@innerbib\@empty
\bibitem [{\citenamefont {Krawczyk}\ and\ \citenamefont
  {Grundler}(2014)}]{Krawczyk2014}%
  \BibitemOpen
  \bibfield  {author} {\bibinfo {author} {\bibfnamefont {M.}~\bibnamefont
  {Krawczyk}}\ and\ \bibinfo {author} {\bibfnamefont {D.}~\bibnamefont
  {Grundler}},\ }\bibfield  {title} {\bibinfo {title} {Review and prospects of
  magnonic crystals and devices with reprogrammable band structure},\
  }\href@noop {} {\bibfield  {journal} {\bibinfo  {journal} {Journal of
  Physics: Condensed Matter}\ }\textbf {\bibinfo {volume} {26}},\ \bibinfo
  {pages} {123202} (\bibinfo {year} {2014})}\BibitemShut {NoStop}%
\bibitem [{\citenamefont {Chumak}\ \emph {et~al.}(2022)\citenamefont {Chumak},
  \citenamefont {Kabos}, \citenamefont {Wu}, \citenamefont {Abert},
  \citenamefont {Adelmann}, \citenamefont {Adeyeye}, \citenamefont
  {\AA{}kerman}, \citenamefont {Aliev}, \citenamefont {Anane}, \citenamefont
  {Awad}, \citenamefont {Back}, \citenamefont {Barman}, \citenamefont {Bauer},
  \citenamefont {Becherer}, \citenamefont {Beginin}, \citenamefont
  {Bittencourt}, \citenamefont {Blanter}, \citenamefont {Bortolotti},
  \citenamefont {Boventer}, \citenamefont {Bozhko}, \citenamefont {Bunyaev},
  \citenamefont {Carmiggelt}, \citenamefont {Cheenikundil}, \citenamefont
  {Ciubotaru}, \citenamefont {Cotofana}, \citenamefont {Csaba}, \citenamefont
  {Dobrovolskiy}, \citenamefont {Dubs}, \citenamefont {Elyasi}, \citenamefont
  {Fripp}, \citenamefont {Fulara}, \citenamefont {Golovchanskiy}, \citenamefont
  {Gonzalez-Ballestero}, \citenamefont {Graczyk}, \citenamefont {Grundler},
  \citenamefont {Gruszecki}, \citenamefont {Gubbiotti}, \citenamefont
  {Guslienko}, \citenamefont {Haldar}, \citenamefont {Hamdioui}, \citenamefont
  {Hertel}, \citenamefont {Hillebrands}, \citenamefont {Hioki}, \citenamefont
  {Houshang}, \citenamefont {Hu}, \citenamefont {Huebl}, \citenamefont {Huth},
  \citenamefont {Iacocca}, \citenamefont {Jungfleisch}, \citenamefont
  {Kakazei}, \citenamefont {Khitun}, \citenamefont {Khymyn}, \citenamefont
  {Kikkawa}, \citenamefont {Kl\"{a}ui}, \citenamefont {Klein}, \citenamefont
  {K\l{l}os}, \citenamefont {Knauer}, \citenamefont {Koraltan}, \citenamefont
  {Kostylev}, \citenamefont {Krawczyk}, \citenamefont {Krivorotov},
  \citenamefont {Kruglyak}, \citenamefont {Lachance-Quirion}, \citenamefont
  {Ladak}, \citenamefont {Lebrun}, \citenamefont {Li}, \citenamefont {Lindner},
  \citenamefont {Mac\^{e}do}, \citenamefont {Mayr}, \citenamefont {Melkov},
  \citenamefont {Mieszczak}, \citenamefont {Nakamura}, \citenamefont {Nembach},
  \citenamefont {Nikitin}, \citenamefont {Nikitov}, \citenamefont {Novosad},
  \citenamefont {Ot\'{a}lora}, \citenamefont {Otani}, \citenamefont {Papp},
  \citenamefont {Pigeau}, \citenamefont {Pirro}, \citenamefont {Porod},
  \citenamefont {Porrati}, \citenamefont {Qin}, \citenamefont {Rana},
  \citenamefont {Reimann}, \citenamefont {Riente}, \citenamefont
  {Romero-Isart}, \citenamefont {Ross}, \citenamefont {Sadovnikov},
  \citenamefont {Safin}, \citenamefont {Saitoh}, \citenamefont {Schmidt},
  \citenamefont {Schultheiss}, \citenamefont {Schultheiss}, \citenamefont
  {Serga}, \citenamefont {Sharma}, \citenamefont {Shaw}, \citenamefont {Suess},
  \citenamefont {Surzhenko}, \citenamefont {Szulc}, \citenamefont {Taniguchi},
  \citenamefont {Urb\'{a}nek}, \citenamefont {Usami}, \citenamefont {Ustinov},
  \citenamefont {van~der Sar}, \citenamefont {van Dijken}, \citenamefont
  {Vasyuchka}, \citenamefont {Verba}, \citenamefont {Kusminskiy}, \citenamefont
  {Wang}, \citenamefont {Weides}, \citenamefont {Weiler}, \citenamefont
  {Wintz}, \citenamefont {Wolski},\ and\ \citenamefont {Zhang}}]{Chumak2022}%
  \BibitemOpen
  \bibfield  {author} {\bibinfo {author} {\bibfnamefont {A.~V.}\ \bibnamefont
  {Chumak}}, \bibinfo {author} {\bibfnamefont {P.}~\bibnamefont {Kabos}},
  \bibinfo {author} {\bibfnamefont {M.}~\bibnamefont {Wu}}, \bibinfo {author}
  {\bibfnamefont {C.}~\bibnamefont {Abert}}, \bibinfo {author} {\bibfnamefont
  {C.}~\bibnamefont {Adelmann}}, \bibinfo {author} {\bibfnamefont {A.~O.}\
  \bibnamefont {Adeyeye}}, \bibinfo {author} {\bibfnamefont {J.}~\bibnamefont
  {\AA{}kerman}}, \bibinfo {author} {\bibfnamefont {F.~G.}\ \bibnamefont
  {Aliev}}, \bibinfo {author} {\bibfnamefont {A.}~\bibnamefont {Anane}},
  \bibinfo {author} {\bibfnamefont {A.}~\bibnamefont {Awad}}, \bibinfo {author}
  {\bibfnamefont {C.~H.}\ \bibnamefont {Back}}, \bibinfo {author}
  {\bibfnamefont {A.}~\bibnamefont {Barman}}, \bibinfo {author} {\bibfnamefont
  {G.~E.~W.}\ \bibnamefont {Bauer}}, \bibinfo {author} {\bibfnamefont
  {M.}~\bibnamefont {Becherer}}, \bibinfo {author} {\bibfnamefont {E.~N.}\
  \bibnamefont {Beginin}}, \bibinfo {author} {\bibfnamefont {V.~A. S.~V.}\
  \bibnamefont {Bittencourt}}, \bibinfo {author} {\bibfnamefont {Y.~M.}\
  \bibnamefont {Blanter}}, \bibinfo {author} {\bibfnamefont {P.}~\bibnamefont
  {Bortolotti}}, \bibinfo {author} {\bibfnamefont {I.}~\bibnamefont
  {Boventer}}, \bibinfo {author} {\bibfnamefont {D.~A.}\ \bibnamefont
  {Bozhko}}, \bibinfo {author} {\bibfnamefont {S.~A.}\ \bibnamefont {Bunyaev}},
  \bibinfo {author} {\bibfnamefont {J.~J.}\ \bibnamefont {Carmiggelt}},
  \bibinfo {author} {\bibfnamefont {R.~R.}\ \bibnamefont {Cheenikundil}},
  \bibinfo {author} {\bibfnamefont {F.}~\bibnamefont {Ciubotaru}}, \bibinfo
  {author} {\bibfnamefont {S.}~\bibnamefont {Cotofana}}, \bibinfo {author}
  {\bibfnamefont {G.}~\bibnamefont {Csaba}}, \bibinfo {author} {\bibfnamefont
  {O.~V.}\ \bibnamefont {Dobrovolskiy}}, \bibinfo {author} {\bibfnamefont
  {C.}~\bibnamefont {Dubs}}, \bibinfo {author} {\bibfnamefont {M.}~\bibnamefont
  {Elyasi}}, \bibinfo {author} {\bibfnamefont {K.~G.}\ \bibnamefont {Fripp}},
  \bibinfo {author} {\bibfnamefont {H.}~\bibnamefont {Fulara}}, \bibinfo
  {author} {\bibfnamefont {I.~A.}\ \bibnamefont {Golovchanskiy}}, \bibinfo
  {author} {\bibfnamefont {C.}~\bibnamefont {Gonzalez-Ballestero}}, \bibinfo
  {author} {\bibfnamefont {P.}~\bibnamefont {Graczyk}}, \bibinfo {author}
  {\bibfnamefont {D.}~\bibnamefont {Grundler}}, \bibinfo {author}
  {\bibfnamefont {P.}~\bibnamefont {Gruszecki}}, \bibinfo {author}
  {\bibfnamefont {G.}~\bibnamefont {Gubbiotti}}, \bibinfo {author}
  {\bibfnamefont {K.}~\bibnamefont {Guslienko}}, \bibinfo {author}
  {\bibfnamefont {A.}~\bibnamefont {Haldar}}, \bibinfo {author} {\bibfnamefont
  {S.}~\bibnamefont {Hamdioui}}, \bibinfo {author} {\bibfnamefont
  {R.}~\bibnamefont {Hertel}}, \bibinfo {author} {\bibfnamefont
  {B.}~\bibnamefont {Hillebrands}}, \bibinfo {author} {\bibfnamefont
  {T.}~\bibnamefont {Hioki}}, \bibinfo {author} {\bibfnamefont
  {A.}~\bibnamefont {Houshang}}, \bibinfo {author} {\bibfnamefont {C.-M.}\
  \bibnamefont {Hu}}, \bibinfo {author} {\bibfnamefont {H.}~\bibnamefont
  {Huebl}}, \bibinfo {author} {\bibfnamefont {M.}~\bibnamefont {Huth}},
  \bibinfo {author} {\bibfnamefont {E.}~\bibnamefont {Iacocca}}, \bibinfo
  {author} {\bibfnamefont {M.~B.}\ \bibnamefont {Jungfleisch}}, \bibinfo
  {author} {\bibfnamefont {G.~N.}\ \bibnamefont {Kakazei}}, \bibinfo {author}
  {\bibfnamefont {A.}~\bibnamefont {Khitun}}, \bibinfo {author} {\bibfnamefont
  {R.}~\bibnamefont {Khymyn}}, \bibinfo {author} {\bibfnamefont
  {T.}~\bibnamefont {Kikkawa}}, \bibinfo {author} {\bibfnamefont
  {M.}~\bibnamefont {Kl\"{a}ui}}, \bibinfo {author} {\bibfnamefont
  {O.}~\bibnamefont {Klein}}, \bibinfo {author} {\bibfnamefont {J.~W.}\
  \bibnamefont {K\l{l}os}}, \bibinfo {author} {\bibfnamefont {S.}~\bibnamefont
  {Knauer}}, \bibinfo {author} {\bibfnamefont {S.}~\bibnamefont {Koraltan}},
  \bibinfo {author} {\bibfnamefont {M.}~\bibnamefont {Kostylev}}, \bibinfo
  {author} {\bibfnamefont {M.}~\bibnamefont {Krawczyk}}, \bibinfo {author}
  {\bibfnamefont {I.~N.}\ \bibnamefont {Krivorotov}}, \bibinfo {author}
  {\bibfnamefont {V.~V.}\ \bibnamefont {Kruglyak}}, \bibinfo {author}
  {\bibfnamefont {D.}~\bibnamefont {Lachance-Quirion}}, \bibinfo {author}
  {\bibfnamefont {S.}~\bibnamefont {Ladak}}, \bibinfo {author} {\bibfnamefont
  {R.}~\bibnamefont {Lebrun}}, \bibinfo {author} {\bibfnamefont
  {Y.}~\bibnamefont {Li}}, \bibinfo {author} {\bibfnamefont {M.}~\bibnamefont
  {Lindner}}, \bibinfo {author} {\bibfnamefont {R.}~\bibnamefont {Mac\^{e}do}},
  \bibinfo {author} {\bibfnamefont {S.}~\bibnamefont {Mayr}}, \bibinfo {author}
  {\bibfnamefont {G.~A.}\ \bibnamefont {Melkov}}, \bibinfo {author}
  {\bibfnamefont {S.}~\bibnamefont {Mieszczak}}, \bibinfo {author}
  {\bibfnamefont {Y.}~\bibnamefont {Nakamura}}, \bibinfo {author}
  {\bibfnamefont {H.~T.}\ \bibnamefont {Nembach}}, \bibinfo {author}
  {\bibfnamefont {A.~A.}\ \bibnamefont {Nikitin}}, \bibinfo {author}
  {\bibfnamefont {S.~A.}\ \bibnamefont {Nikitov}}, \bibinfo {author}
  {\bibfnamefont {V.}~\bibnamefont {Novosad}}, \bibinfo {author} {\bibfnamefont
  {J.~A.}\ \bibnamefont {Ot\'{a}lora}}, \bibinfo {author} {\bibfnamefont
  {Y.}~\bibnamefont {Otani}}, \bibinfo {author} {\bibfnamefont
  {A.}~\bibnamefont {Papp}}, \bibinfo {author} {\bibfnamefont {B.}~\bibnamefont
  {Pigeau}}, \bibinfo {author} {\bibfnamefont {P.}~\bibnamefont {Pirro}},
  \bibinfo {author} {\bibfnamefont {W.}~\bibnamefont {Porod}}, \bibinfo
  {author} {\bibfnamefont {F.}~\bibnamefont {Porrati}}, \bibinfo {author}
  {\bibfnamefont {H.}~\bibnamefont {Qin}}, \bibinfo {author} {\bibfnamefont
  {B.}~\bibnamefont {Rana}}, \bibinfo {author} {\bibfnamefont {T.}~\bibnamefont
  {Reimann}}, \bibinfo {author} {\bibfnamefont {F.}~\bibnamefont {Riente}},
  \bibinfo {author} {\bibfnamefont {O.}~\bibnamefont {Romero-Isart}}, \bibinfo
  {author} {\bibfnamefont {A.}~\bibnamefont {Ross}}, \bibinfo {author}
  {\bibfnamefont {A.~V.}\ \bibnamefont {Sadovnikov}}, \bibinfo {author}
  {\bibfnamefont {A.~R.}\ \bibnamefont {Safin}}, \bibinfo {author}
  {\bibfnamefont {E.}~\bibnamefont {Saitoh}}, \bibinfo {author} {\bibfnamefont
  {G.}~\bibnamefont {Schmidt}}, \bibinfo {author} {\bibfnamefont
  {H.}~\bibnamefont {Schultheiss}}, \bibinfo {author} {\bibfnamefont
  {K.}~\bibnamefont {Schultheiss}}, \bibinfo {author} {\bibfnamefont {A.~A.}\
  \bibnamefont {Serga}}, \bibinfo {author} {\bibfnamefont {S.}~\bibnamefont
  {Sharma}}, \bibinfo {author} {\bibfnamefont {J.~M.}\ \bibnamefont {Shaw}},
  \bibinfo {author} {\bibfnamefont {D.}~\bibnamefont {Suess}}, \bibinfo
  {author} {\bibfnamefont {O.}~\bibnamefont {Surzhenko}}, \bibinfo {author}
  {\bibfnamefont {K.}~\bibnamefont {Szulc}}, \bibinfo {author} {\bibfnamefont
  {T.}~\bibnamefont {Taniguchi}}, \bibinfo {author} {\bibfnamefont
  {M.}~\bibnamefont {Urb\'{a}nek}}, \bibinfo {author} {\bibfnamefont
  {K.}~\bibnamefont {Usami}}, \bibinfo {author} {\bibfnamefont {A.~B.}\
  \bibnamefont {Ustinov}}, \bibinfo {author} {\bibfnamefont {T.}~\bibnamefont
  {van~der Sar}}, \bibinfo {author} {\bibfnamefont {S.}~\bibnamefont {van
  Dijken}}, \bibinfo {author} {\bibfnamefont {V.~I.}\ \bibnamefont
  {Vasyuchka}}, \bibinfo {author} {\bibfnamefont {R.}~\bibnamefont {Verba}},
  \bibinfo {author} {\bibfnamefont {S.~V.}\ \bibnamefont {Kusminskiy}},
  \bibinfo {author} {\bibfnamefont {Q.}~\bibnamefont {Wang}}, \bibinfo {author}
  {\bibfnamefont {M.}~\bibnamefont {Weides}}, \bibinfo {author} {\bibfnamefont
  {M.}~\bibnamefont {Weiler}}, \bibinfo {author} {\bibfnamefont
  {S.}~\bibnamefont {Wintz}}, \bibinfo {author} {\bibfnamefont {S.~P.}\
  \bibnamefont {Wolski}},\ and\ \bibinfo {author} {\bibfnamefont
  {X.}~\bibnamefont {Zhang}},\ }\bibfield  {title} {\bibinfo {title} {Advances
  in magnetics roadmap on spin-wave computing},\ }\href
  {https://doi.org/10.1109/TMAG.2022.3149664} {\bibfield  {journal} {\bibinfo
  {journal} {IEEE Transactions on Magnetics}\ }\textbf {\bibinfo {volume}
  {58}},\ \bibinfo {pages} {1} (\bibinfo {year} {2022})}\BibitemShut {NoStop}%
\bibitem [{\citenamefont {Nikitov}\ \emph {et~al.}(2001)\citenamefont
  {Nikitov}, \citenamefont {Tailhades},\ and\ \citenamefont
  {Tsai}}]{Nikitov2001}%
  \BibitemOpen
  \bibfield  {author} {\bibinfo {author} {\bibfnamefont {S.}~\bibnamefont
  {Nikitov}}, \bibinfo {author} {\bibfnamefont {P.}~\bibnamefont {Tailhades}},\
  and\ \bibinfo {author} {\bibfnamefont {C.}~\bibnamefont {Tsai}},\ }\bibfield
  {title} {\bibinfo {title} {Spin waves in periodic magnetic
  structures-magnonic crystals},\ }\href@noop {} {\bibfield  {journal}
  {\bibinfo  {journal} {Journal of Magnetism and Magnetic Materials}\ }\textbf
  {\bibinfo {volume} {236}},\ \bibinfo {pages} {320 } (\bibinfo {year}
  {2001})}\BibitemShut {NoStop}%
\bibitem [{\citenamefont {Neusser}\ and\ \citenamefont
  {Grundler}(2009)}]{Neusser2009}%
  \BibitemOpen
  \bibfield  {author} {\bibinfo {author} {\bibfnamefont {S.}~\bibnamefont
  {Neusser}}\ and\ \bibinfo {author} {\bibfnamefont {D.}~\bibnamefont
  {Grundler}},\ }\bibfield  {title} {\bibinfo {title} {Magnonics: Spin waves on
  the nanoscale},\ }\href@noop {} {\bibfield  {journal} {\bibinfo  {journal}
  {Advanced Materials}\ }\textbf {\bibinfo {volume} {21}},\ \bibinfo {pages}
  {2927} (\bibinfo {year} {2009})}\BibitemShut {NoStop}%
\bibitem [{\citenamefont {Wang}\ \emph {et~al.}(2010)\citenamefont {Wang},
  \citenamefont {Zhang}, \citenamefont {Lim}, \citenamefont {Ng}, \citenamefont
  {Kuok}, \citenamefont {Jain},\ and\ \citenamefont {Adeyeye}}]{Wang2010}%
  \BibitemOpen
  \bibfield  {author} {\bibinfo {author} {\bibfnamefont {Z.}~\bibnamefont
  {Wang}}, \bibinfo {author} {\bibfnamefont {V.}~\bibnamefont {Zhang}},
  \bibinfo {author} {\bibfnamefont {H.}~\bibnamefont {Lim}}, \bibinfo {author}
  {\bibfnamefont {S.}~\bibnamefont {Ng}}, \bibinfo {author} {\bibfnamefont
  {M.}~\bibnamefont {Kuok}}, \bibinfo {author} {\bibfnamefont {S.}~\bibnamefont
  {Jain}},\ and\ \bibinfo {author} {\bibfnamefont {A.}~\bibnamefont
  {Adeyeye}},\ }\bibfield  {title} {\bibinfo {title} {Nanostructured magnonic
  crystals with size-tunable bandgaps},\ }\href
  {http://pubs.acs.org/doi/abs/10.1021/nn901171u} {\bibfield  {journal}
  {\bibinfo  {journal} {ACS Nano}\ }\textbf {\bibinfo {volume} {4}},\ \bibinfo
  {pages} {643} (\bibinfo {year} {2010})}\BibitemShut {NoStop}%
\bibitem [{\citenamefont {Kumar}\ \emph {et~al.}(2014)\citenamefont {Kumar},
  \citenamefont {K\l{}os}, \citenamefont {Krawczyk},\ and\ \citenamefont
  {Barman}}]{Kumar2014}%
  \BibitemOpen
  \bibfield  {author} {\bibinfo {author} {\bibfnamefont {D.}~\bibnamefont
  {Kumar}}, \bibinfo {author} {\bibfnamefont {J.~W.}\ \bibnamefont {K\l{}os}},
  \bibinfo {author} {\bibfnamefont {M.}~\bibnamefont {Krawczyk}},\ and\
  \bibinfo {author} {\bibfnamefont {A.}~\bibnamefont {Barman}},\ }\bibfield
  {title} {\bibinfo {title} {Magnonic band structure, complete bandgap, and
  collective spin wave excitation in nanoscale two-dimensional magnonic
  crystals},\ }\href {https://doi.org/10.1063/1.4862911} {\bibfield  {journal}
  {\bibinfo  {journal} {Journal of Applied Physics}\ }\textbf {\bibinfo
  {volume} {115}},\ \bibinfo {pages} {043917} (\bibinfo {year}
  {2014})}\BibitemShut {NoStop}%
\bibitem [{\citenamefont {Gubbiotti}\ \emph {et~al.}(2018)\citenamefont
  {Gubbiotti}, \citenamefont {Zhou}, \citenamefont {Haghshenasfard},
  \citenamefont {Cottam},\ and\ \citenamefont {Adeyeye}}]{Gubbiotti2018}%
  \BibitemOpen
  \bibfield  {author} {\bibinfo {author} {\bibfnamefont {G.}~\bibnamefont
  {Gubbiotti}}, \bibinfo {author} {\bibfnamefont {X.}~\bibnamefont {Zhou}},
  \bibinfo {author} {\bibfnamefont {Z.}~\bibnamefont {Haghshenasfard}},
  \bibinfo {author} {\bibfnamefont {M.~G.}\ \bibnamefont {Cottam}},\ and\
  \bibinfo {author} {\bibfnamefont {A.~O.}\ \bibnamefont {Adeyeye}},\
  }\bibfield  {title} {\bibinfo {title} {Reprogrammable magnonic band structure
  of layered permalloy/cu/permalloy nanowires},\ }\href@noop {} {\bibfield
  {journal} {\bibinfo  {journal} {Phys. Rev. B}\ }\textbf {\bibinfo {volume}
  {97}},\ \bibinfo {pages} {134428} (\bibinfo {year} {2018})}\BibitemShut
  {NoStop}%
\bibitem [{\citenamefont {Lisiecki}\ \emph {et~al.}(2019)\citenamefont
  {Lisiecki}, \citenamefont {Rych\l{}y}, \citenamefont
  {Ku\ifmmode~\acute{s}\else \'{s}\fi{}wik}, \citenamefont
  {G\l{}owi\ifmmode~\acute{n}\else \'{n}\fi{}ski}, \citenamefont {K\l{}os},
  \citenamefont {Gro\ss{}}, \citenamefont {Tr\"ager}, \citenamefont {Bykova},
  \citenamefont {Weigand}, \citenamefont {Zelent}, \citenamefont {Goering},
  \citenamefont {Sch\"utz}, \citenamefont {Krawczyk}, \citenamefont
  {Stobiecki}, \citenamefont {Dubowik},\ and\ \citenamefont
  {Gr\"afe}}]{Lisiecki2019}%
  \BibitemOpen
  \bibfield  {author} {\bibinfo {author} {\bibfnamefont {F.}~\bibnamefont
  {Lisiecki}}, \bibinfo {author} {\bibfnamefont {J.}~\bibnamefont {Rych\l{}y}},
  \bibinfo {author} {\bibfnamefont {P.}~\bibnamefont {Ku\ifmmode~\acute{s}\else
  \'{s}\fi{}wik}}, \bibinfo {author} {\bibfnamefont {H.}~\bibnamefont
  {G\l{}owi\ifmmode~\acute{n}\else \'{n}\fi{}ski}}, \bibinfo {author}
  {\bibfnamefont {J.~W.}\ \bibnamefont {K\l{}os}}, \bibinfo {author}
  {\bibfnamefont {F.}~\bibnamefont {Gro\ss{}}}, \bibinfo {author}
  {\bibfnamefont {N.}~\bibnamefont {Tr\"ager}}, \bibinfo {author}
  {\bibfnamefont {I.}~\bibnamefont {Bykova}}, \bibinfo {author} {\bibfnamefont
  {M.}~\bibnamefont {Weigand}}, \bibinfo {author} {\bibfnamefont
  {M.}~\bibnamefont {Zelent}}, \bibinfo {author} {\bibfnamefont {E.~J.}\
  \bibnamefont {Goering}}, \bibinfo {author} {\bibfnamefont {G.}~\bibnamefont
  {Sch\"utz}}, \bibinfo {author} {\bibfnamefont {M.}~\bibnamefont {Krawczyk}},
  \bibinfo {author} {\bibfnamefont {F.}~\bibnamefont {Stobiecki}}, \bibinfo
  {author} {\bibfnamefont {J.}~\bibnamefont {Dubowik}},\ and\ \bibinfo {author}
  {\bibfnamefont {J.}~\bibnamefont {Gr\"afe}},\ }\bibfield  {title} {\bibinfo
  {title} {Magnons in a quasicrystal: Propagation, extinction, and localization
  of spin waves in fibonacci structures},\ }\href@noop {} {\bibfield  {journal}
  {\bibinfo  {journal} {Phys. Rev. Applied}\ }\textbf {\bibinfo {volume}
  {11}},\ \bibinfo {pages} {054061} (\bibinfo {year} {2019})}\BibitemShut
  {NoStop}%
\bibitem [{\citenamefont {Szulc}\ \emph {et~al.}(2022)\citenamefont {Szulc},
  \citenamefont {Tacchi}, \citenamefont {Hierro-Rogr\'{i}guez}, \citenamefont
  {D\'{i}az}, \citenamefont {Gruszecki}, \citenamefont {Graczyk}, \citenamefont
  {Quir\'{o}s}, \citenamefont {Mark\'{o}}, \citenamefont {Mart\'{i}n},
  \citenamefont {V\'{e}lez}, \citenamefont {Schmool}, \citenamefont {Carlotti},
  \citenamefont {Krawczyk},\ and\ \citenamefont {\'{a}lvarez
  Prado}}]{Szulc2022}%
  \BibitemOpen
  \bibfield  {author} {\bibinfo {author} {\bibfnamefont {K.}~\bibnamefont
  {Szulc}}, \bibinfo {author} {\bibfnamefont {S.}~\bibnamefont {Tacchi}},
  \bibinfo {author} {\bibfnamefont {A.}~\bibnamefont {Hierro-Rogr\'{i}guez}},
  \bibinfo {author} {\bibfnamefont {J.}~\bibnamefont {D\'{i}az}}, \bibinfo
  {author} {\bibfnamefont {P.}~\bibnamefont {Gruszecki}}, \bibinfo {author}
  {\bibfnamefont {P.}~\bibnamefont {Graczyk}}, \bibinfo {author} {\bibfnamefont
  {C.}~\bibnamefont {Quir\'{o}s}}, \bibinfo {author} {\bibfnamefont
  {D.}~\bibnamefont {Mark\'{o}}}, \bibinfo {author} {\bibfnamefont {J.~I.}\
  \bibnamefont {Mart\'{i}n}}, \bibinfo {author} {\bibfnamefont
  {M.}~\bibnamefont {V\'{e}lez}}, \bibinfo {author} {\bibfnamefont {D.~S.}\
  \bibnamefont {Schmool}}, \bibinfo {author} {\bibfnamefont {G.}~\bibnamefont
  {Carlotti}}, \bibinfo {author} {\bibfnamefont {M.}~\bibnamefont {Krawczyk}},\
  and\ \bibinfo {author} {\bibfnamefont {L.}~\bibnamefont {\'{a}lvarez
  Prado}},\ }\bibfield  {title} {\bibinfo {title} {Reconfigurable magnonic
  crystals based on imprinted magnetization textures in hard and soft
  dipolar-coupled bilayers},\ }\href
  {https://pubs.acs.org/doi/10.1021/acsnano.2c04256} {\bibfield  {journal}
  {\bibinfo  {journal} {ACS Nano}\ }\textbf {\bibinfo {volume} {16}},\ \bibinfo
  {pages} {14168} (\bibinfo {year} {2022})}\BibitemShut {NoStop}%
\bibitem [{\citenamefont {Negrello}\ \emph {et~al.}(2022)\citenamefont
  {Negrello}, \citenamefont {Montoncello}, \citenamefont {Kaffash},
  \citenamefont {Jungfleisch},\ and\ \citenamefont {Gubbiotti}}]{Negrello2022}%
  \BibitemOpen
  \bibfield  {author} {\bibinfo {author} {\bibfnamefont {R.}~\bibnamefont
  {Negrello}}, \bibinfo {author} {\bibfnamefont {F.}~\bibnamefont
  {Montoncello}}, \bibinfo {author} {\bibfnamefont {M.~T.}\ \bibnamefont
  {Kaffash}}, \bibinfo {author} {\bibfnamefont {M.~B.}\ \bibnamefont
  {Jungfleisch}},\ and\ \bibinfo {author} {\bibfnamefont {G.}~\bibnamefont
  {Gubbiotti}},\ }\bibfield  {title} {\bibinfo {title} {Dynamic coupling and
  spin-wave dispersions in a magnetic hybrid system made of an artificial
  spin-ice structure and an extended nife underlayer},\ }\href
  {https://doi.org/10.1063/5.0102571} {\bibfield  {journal} {\bibinfo
  {journal} {APL Materials}\ }\textbf {\bibinfo {volume} {10}},\ \bibinfo
  {pages} {091115} (\bibinfo {year} {2022})}\BibitemShut {NoStop}%
\bibitem [{\citenamefont {Roxburgh}\ and\ \citenamefont
  {Iacocca}(2024)}]{Roxburgh2024}%
  \BibitemOpen
  \bibfield  {author} {\bibinfo {author} {\bibfnamefont {A.}~\bibnamefont
  {Roxburgh}}\ and\ \bibinfo {author} {\bibfnamefont {E.}~\bibnamefont
  {Iacocca}},\ }\bibfield  {title} {\bibinfo {title} {Nano-magnonic crystals by
  periodic modulation of magnetic parameters},\ }\bibfield  {journal} {\bibinfo
   {journal} {Magnetochemistry}\ }\textbf {\bibinfo {volume} {10}},\ \href
  {https://doi.org/10.3390/magnetochemistry10030014}
  {10.3390/magnetochemistry10030014} (\bibinfo {year} {2024})\BibitemShut
  {NoStop}%
\bibitem [{\citenamefont {Roxburgh}\ \emph {et~al.}(2025)\citenamefont
  {Roxburgh}, \citenamefont {Micaletti}, \citenamefont {Montoncello},\ and\
  \citenamefont {Iacocca}}]{Roxburgh2025}%
  \BibitemOpen
  \bibfield  {author} {\bibinfo {author} {\bibfnamefont {A.}~\bibnamefont
  {Roxburgh}}, \bibinfo {author} {\bibfnamefont {P.}~\bibnamefont {Micaletti}},
  \bibinfo {author} {\bibfnamefont {F.}~\bibnamefont {Montoncello}},\ and\
  \bibinfo {author} {\bibfnamefont {E.}~\bibnamefont {Iacocca}},\ }\bibfield
  {title} {\bibinfo {title} {Generation of spin-wave packets by reconfigurable
  nanomagnonic heterojunctions},\ }\href {https://doi.org/10.1103/bkqp-vjhy}
  {\bibfield  {journal} {\bibinfo  {journal} {Phys. Rev. Appl.}\ }\textbf
  {\bibinfo {volume} {24}},\ \bibinfo {pages} {014047} (\bibinfo {year}
  {2025})}\BibitemShut {NoStop}%
\bibitem [{\citenamefont {Tacchi}\ \emph {et~al.}(2011)\citenamefont {Tacchi},
  \citenamefont {Montoncello}, \citenamefont {Madami}, \citenamefont
  {Gubbiotti}, \citenamefont {Carlotti}, \citenamefont {Giovannini},
  \citenamefont {Zivieri}, \citenamefont {Nizzoli}, \citenamefont {Jain},
  \citenamefont {Adeyeye},\ and\ \citenamefont {Singh}}]{Tacchi2011}%
  \BibitemOpen
  \bibfield  {author} {\bibinfo {author} {\bibfnamefont {S.}~\bibnamefont
  {Tacchi}}, \bibinfo {author} {\bibfnamefont {F.}~\bibnamefont {Montoncello}},
  \bibinfo {author} {\bibfnamefont {M.}~\bibnamefont {Madami}}, \bibinfo
  {author} {\bibfnamefont {G.}~\bibnamefont {Gubbiotti}}, \bibinfo {author}
  {\bibfnamefont {G.}~\bibnamefont {Carlotti}}, \bibinfo {author}
  {\bibfnamefont {L.}~\bibnamefont {Giovannini}}, \bibinfo {author}
  {\bibfnamefont {R.}~\bibnamefont {Zivieri}}, \bibinfo {author} {\bibfnamefont
  {F.}~\bibnamefont {Nizzoli}}, \bibinfo {author} {\bibfnamefont
  {S.}~\bibnamefont {Jain}}, \bibinfo {author} {\bibfnamefont {A.~O.}\
  \bibnamefont {Adeyeye}},\ and\ \bibinfo {author} {\bibfnamefont
  {N.}~\bibnamefont {Singh}},\ }\bibfield  {title} {\bibinfo {title} {Band
  diagram of spin waves in a two-dimensional magnonic crystal},\ }\href@noop {}
  {\bibfield  {journal} {\bibinfo  {journal} {Phys. Rev. Lett.}\ }\textbf
  {\bibinfo {volume} {107}},\ \bibinfo {pages} {127204} (\bibinfo {year}
  {2011})}\BibitemShut {NoStop}%
\bibitem [{\citenamefont {Neusser}\ \emph {et~al.}(2011)\citenamefont
  {Neusser}, \citenamefont {Duerr}, \citenamefont {Tacchi}, \citenamefont
  {Madami}, \citenamefont {Sokolovskyy}, \citenamefont {Gubbiotti},
  \citenamefont {Krawczyk},\ and\ \citenamefont {Grundler}}]{Neusser2011}%
  \BibitemOpen
  \bibfield  {author} {\bibinfo {author} {\bibfnamefont {S.}~\bibnamefont
  {Neusser}}, \bibinfo {author} {\bibfnamefont {G.}~\bibnamefont {Duerr}},
  \bibinfo {author} {\bibfnamefont {S.}~\bibnamefont {Tacchi}}, \bibinfo
  {author} {\bibfnamefont {M.}~\bibnamefont {Madami}}, \bibinfo {author}
  {\bibfnamefont {M.~L.}\ \bibnamefont {Sokolovskyy}}, \bibinfo {author}
  {\bibfnamefont {G.}~\bibnamefont {Gubbiotti}}, \bibinfo {author}
  {\bibfnamefont {M.}~\bibnamefont {Krawczyk}},\ and\ \bibinfo {author}
  {\bibfnamefont {D.}~\bibnamefont {Grundler}},\ }\bibfield  {title} {\bibinfo
  {title} {Magnonic minibands in antidot lattices with large spin-wave
  propagation velocities},\ }\href@noop {} {\bibfield  {journal} {\bibinfo
  {journal} {Phys. Rev. B}\ }\textbf {\bibinfo {volume} {84}},\ \bibinfo
  {pages} {094454} (\bibinfo {year} {2011})}\BibitemShut {NoStop}%
\bibitem [{\citenamefont {Gliga}\ \emph {et~al.}(2020)\citenamefont {Gliga},
  \citenamefont {Iacocca},\ and\ \citenamefont {Heinonen}}]{Gliga2020}%
  \BibitemOpen
  \bibfield  {author} {\bibinfo {author} {\bibfnamefont {S.}~\bibnamefont
  {Gliga}}, \bibinfo {author} {\bibfnamefont {E.}~\bibnamefont {Iacocca}},\
  and\ \bibinfo {author} {\bibfnamefont {O.~G.}\ \bibnamefont {Heinonen}},\
  }\bibfield  {title} {\bibinfo {title} {Dynamics of reconfigurable artificial
  spin ice: Toward magnonic functional materials},\ }\href
  {https://doi.org/10.1063/1.5142705} {\bibfield  {journal} {\bibinfo
  {journal} {APL Materials}\ }\textbf {\bibinfo {volume} {8}},\ \bibinfo
  {pages} {040911} (\bibinfo {year} {2020})},\ \Eprint
  {https://arxiv.org/abs/https://doi.org/10.1063/1.5142705}
  {https://doi.org/10.1063/1.5142705} \BibitemShut {NoStop}%
\bibitem [{\citenamefont {Mamica}\ \emph {et~al.}(2018)\citenamefont {Mamica},
  \citenamefont {Zhou}, \citenamefont {Adeyeye}, \citenamefont {Krawczyk},\
  and\ \citenamefont {Gubbiotti}}]{Mamica2018}%
  \BibitemOpen
  \bibfield  {author} {\bibinfo {author} {\bibfnamefont {S.}~\bibnamefont
  {Mamica}}, \bibinfo {author} {\bibfnamefont {X.}~\bibnamefont {Zhou}},
  \bibinfo {author} {\bibfnamefont {A.}~\bibnamefont {Adeyeye}}, \bibinfo
  {author} {\bibfnamefont {M.}~\bibnamefont {Krawczyk}},\ and\ \bibinfo
  {author} {\bibfnamefont {G.}~\bibnamefont {Gubbiotti}},\ }\bibfield  {title}
  {\bibinfo {title} {Spin-wave dynamics in artificial anti-spin-ice systems:
  Experimental and theoretical investigations},\ }\href
  {https://doi.org/10.1103/PhysRevB.98.054405} {\bibfield  {journal} {\bibinfo
  {journal} {Phys. Rev. B}\ }\textbf {\bibinfo {volume} {98}},\ \bibinfo
  {pages} {054405} (\bibinfo {year} {2018})}\BibitemShut {NoStop}%
\bibitem [{\citenamefont {Iacocca}\ \emph {et~al.}(2016)\citenamefont
  {Iacocca}, \citenamefont {Gliga}, \citenamefont {Stamps},\ and\ \citenamefont
  {Heinonen}}]{Iacocca2016}%
  \BibitemOpen
  \bibfield  {author} {\bibinfo {author} {\bibfnamefont {E.}~\bibnamefont
  {Iacocca}}, \bibinfo {author} {\bibfnamefont {S.}~\bibnamefont {Gliga}},
  \bibinfo {author} {\bibfnamefont {R.~L.}\ \bibnamefont {Stamps}},\ and\
  \bibinfo {author} {\bibfnamefont {O.}~\bibnamefont {Heinonen}},\ }\bibfield
  {title} {\bibinfo {title} {Reconfigurable wave band structure of an
  artificial square ice},\ }\href {https://doi.org/10.1103/PhysRevB.93.134420}
  {\bibfield  {journal} {\bibinfo  {journal} {Phys. Rev. B}\ }\textbf {\bibinfo
  {volume} {93}},\ \bibinfo {pages} {134420} (\bibinfo {year}
  {2016})}\BibitemShut {NoStop}%
\bibitem [{\citenamefont {Wysin}\ \emph {et~al.}(2013)\citenamefont {Wysin},
  \citenamefont {Moura-Melo}, \citenamefont {M\`{o}l},\ and\ \citenamefont
  {Pereira}}]{Wysin2013}%
  \BibitemOpen
  \bibfield  {author} {\bibinfo {author} {\bibfnamefont {G.~M.}\ \bibnamefont
  {Wysin}}, \bibinfo {author} {\bibfnamefont {W.~A.}\ \bibnamefont
  {Moura-Melo}}, \bibinfo {author} {\bibfnamefont {L.~A.~S.}\ \bibnamefont
  {M\`{o}l}},\ and\ \bibinfo {author} {\bibfnamefont {A.~R.}\ \bibnamefont
  {Pereira}},\ }\bibfield  {title} {\bibinfo {title} {Dynamics and hysteresis
  in square lattice artificial spin ice},\ }\href
  {https://doi.org/10.1088/1367-2630/15/4/045029} {\bibfield  {journal}
  {\bibinfo  {journal} {New Journal of Physics}\ }\textbf {\bibinfo {volume}
  {15}},\ \bibinfo {pages} {045029} (\bibinfo {year} {2013})}\BibitemShut
  {NoStop}%
\bibitem [{\citenamefont {Lasnier}\ and\ \citenamefont
  {Wysin}(2020)}]{Lasnier2020}%
  \BibitemOpen
  \bibfield  {author} {\bibinfo {author} {\bibfnamefont {T.~D.}\ \bibnamefont
  {Lasnier}}\ and\ \bibinfo {author} {\bibfnamefont {G.~M.}\ \bibnamefont
  {Wysin}},\ }\bibfield  {title} {\bibinfo {title} {Magnetic oscillation modes
  in square-lattice artificial spin ice},\ }\href
  {https://doi.org/10.1103/PhysRevB.101.224428} {\bibfield  {journal} {\bibinfo
   {journal} {Phys. Rev. B}\ }\textbf {\bibinfo {volume} {101}},\ \bibinfo
  {pages} {224428} (\bibinfo {year} {2020})}\BibitemShut {NoStop}%
\bibitem [{\citenamefont {Heyderman}\ and\ \citenamefont
  {Stamps}(2013)}]{Heyderman2013}%
  \BibitemOpen
  \bibfield  {author} {\bibinfo {author} {\bibfnamefont {L.~J.}\ \bibnamefont
  {Heyderman}}\ and\ \bibinfo {author} {\bibfnamefont {R.~L.}\ \bibnamefont
  {Stamps}},\ }\bibfield  {title} {\bibinfo {title} {Artificial ferroic
  systems: Novel functionality from structure, interactions and dynamics},\
  }\href@noop {} {\bibfield  {journal} {\bibinfo  {journal} {Journal of
  Physics: Condensed Matter}\ }\textbf {\bibinfo {volume} {25}},\ \bibinfo
  {pages} {363201} (\bibinfo {year} {2013})}\BibitemShut {NoStop}%
\bibitem [{\citenamefont {Skj{\ae}rv{\o}}\ \emph {et~al.}(2020)\citenamefont
  {Skj{\ae}rv{\o}}, \citenamefont {Marrows}, \citenamefont {Stamps},\ and\
  \citenamefont {Heyderman}}]{Skjaervo2020}%
  \BibitemOpen
  \bibfield  {author} {\bibinfo {author} {\bibfnamefont {S.~H.}\ \bibnamefont
  {Skj{\ae}rv{\o}}}, \bibinfo {author} {\bibfnamefont {C.~H.}\ \bibnamefont
  {Marrows}}, \bibinfo {author} {\bibfnamefont {R.~L.}\ \bibnamefont
  {Stamps}},\ and\ \bibinfo {author} {\bibfnamefont {L.~J.}\ \bibnamefont
  {Heyderman}},\ }\bibfield  {title} {\bibinfo {title} {Advances in artificial
  spin ice},\ }\href@noop {} {\bibfield  {journal} {\bibinfo  {journal} {Nat.
  Rev. Phys.}\ }\textbf {\bibinfo {volume} {2}},\ \bibinfo {pages} {13}
  (\bibinfo {year} {2020})}\BibitemShut {NoStop}%
\bibitem [{\citenamefont {Lendinez}\ and\ \citenamefont
  {Jungfleisch}(2019)}]{Lendinez2019}%
  \BibitemOpen
  \bibfield  {author} {\bibinfo {author} {\bibfnamefont {S.}~\bibnamefont
  {Lendinez}}\ and\ \bibinfo {author} {\bibfnamefont {M.~B.}\ \bibnamefont
  {Jungfleisch}},\ }\bibfield  {title} {\bibinfo {title} {Magnetization
  dynamics in artificial spin ice},\ }\href@noop {} {\bibfield  {journal}
  {\bibinfo  {journal} {J. Phys.: Condens. Matter}\ }\textbf {\bibinfo {volume}
  {32}},\ \bibinfo {pages} {013001} (\bibinfo {year} {2019})}\BibitemShut
  {NoStop}%
\bibitem [{\citenamefont {Gliga}\ \emph {et~al.}(2013)\citenamefont {Gliga},
  \citenamefont {K\'akay}, \citenamefont {Hertel},\ and\ \citenamefont
  {Heinonen}}]{Gliga2013}%
  \BibitemOpen
  \bibfield  {author} {\bibinfo {author} {\bibfnamefont {S.}~\bibnamefont
  {Gliga}}, \bibinfo {author} {\bibfnamefont {A.}~\bibnamefont {K\'akay}},
  \bibinfo {author} {\bibfnamefont {R.}~\bibnamefont {Hertel}},\ and\ \bibinfo
  {author} {\bibfnamefont {O.~G.}\ \bibnamefont {Heinonen}},\ }\bibfield
  {title} {\bibinfo {title} {Spectral analysis of topological defects in an
  artificial spin-ice lattice},\ }\href
  {https://doi.org/10.1103/PhysRevLett.110.117205} {\bibfield  {journal}
  {\bibinfo  {journal} {Phys. Rev. Lett.}\ }\textbf {\bibinfo {volume} {110}},\
  \bibinfo {pages} {117205} (\bibinfo {year} {2013})}\BibitemShut {NoStop}%
\bibitem [{\citenamefont {Jungfleisch}\ \emph {et~al.}(2016)\citenamefont
  {Jungfleisch}, \citenamefont {Zhang}, \citenamefont {Iacocca}, \citenamefont
  {Sklenar}, \citenamefont {Ding}, \citenamefont {Jiang}, \citenamefont
  {Zhang}, \citenamefont {Pearson}, \citenamefont {Novosad}, \citenamefont
  {Ketterson}, \citenamefont {Heinonen},\ and\ \citenamefont
  {Hoffmann}}]{Jungfleisch2016}%
  \BibitemOpen
  \bibfield  {author} {\bibinfo {author} {\bibfnamefont {M.~B.}\ \bibnamefont
  {Jungfleisch}}, \bibinfo {author} {\bibfnamefont {W.}~\bibnamefont {Zhang}},
  \bibinfo {author} {\bibfnamefont {E.}~\bibnamefont {Iacocca}}, \bibinfo
  {author} {\bibfnamefont {J.}~\bibnamefont {Sklenar}}, \bibinfo {author}
  {\bibfnamefont {J.}~\bibnamefont {Ding}}, \bibinfo {author} {\bibfnamefont
  {W.}~\bibnamefont {Jiang}}, \bibinfo {author} {\bibfnamefont
  {S.}~\bibnamefont {Zhang}}, \bibinfo {author} {\bibfnamefont {J.~E.}\
  \bibnamefont {Pearson}}, \bibinfo {author} {\bibfnamefont {V.}~\bibnamefont
  {Novosad}}, \bibinfo {author} {\bibfnamefont {J.~B.}\ \bibnamefont
  {Ketterson}}, \bibinfo {author} {\bibfnamefont {O.}~\bibnamefont
  {Heinonen}},\ and\ \bibinfo {author} {\bibfnamefont {A.}~\bibnamefont
  {Hoffmann}},\ }\bibfield  {title} {\bibinfo {title} {Dynamic response of an
  artificial square spin ice},\ }\href
  {https://doi.org/10.1103/PhysRevB.93.100401} {\bibfield  {journal} {\bibinfo
  {journal} {Phys. Rev. B}\ }\textbf {\bibinfo {volume} {93}},\ \bibinfo
  {pages} {100401} (\bibinfo {year} {2016})}\BibitemShut {NoStop}%
\bibitem [{\citenamefont {Arroo}\ \emph {et~al.}(2019)\citenamefont {Arroo},
  \citenamefont {Gartside},\ and\ \citenamefont {Branford}}]{Arroo2019}%
  \BibitemOpen
  \bibfield  {author} {\bibinfo {author} {\bibfnamefont {D.~M.}\ \bibnamefont
  {Arroo}}, \bibinfo {author} {\bibfnamefont {J.~C.}\ \bibnamefont
  {Gartside}},\ and\ \bibinfo {author} {\bibfnamefont {W.~R.}\ \bibnamefont
  {Branford}},\ }\bibfield  {title} {\bibinfo {title} {Sculpting the spin-wave
  response of artificial spin ice via microstate selection},\ }\href
  {https://doi.org/10.1103/PhysRevB.100.214425} {\bibfield  {journal} {\bibinfo
   {journal} {Phys. Rev. B}\ }\textbf {\bibinfo {volume} {100}},\ \bibinfo
  {pages} {214425} (\bibinfo {year} {2019})}\BibitemShut {NoStop}%
\bibitem [{\citenamefont {Dion}\ \emph {et~al.}(2019)\citenamefont {Dion},
  \citenamefont {Arroo}, \citenamefont {Yamanoi}, \citenamefont {Kimura},
  \citenamefont {Gartside}, \citenamefont {Cohen}, \citenamefont
  {Kurebayashi},\ and\ \citenamefont {Branford}}]{Dion2019}%
  \BibitemOpen
  \bibfield  {author} {\bibinfo {author} {\bibfnamefont {T.}~\bibnamefont
  {Dion}}, \bibinfo {author} {\bibfnamefont {D.~M.}\ \bibnamefont {Arroo}},
  \bibinfo {author} {\bibfnamefont {K.}~\bibnamefont {Yamanoi}}, \bibinfo
  {author} {\bibfnamefont {T.}~\bibnamefont {Kimura}}, \bibinfo {author}
  {\bibfnamefont {J.~C.}\ \bibnamefont {Gartside}}, \bibinfo {author}
  {\bibfnamefont {L.~F.}\ \bibnamefont {Cohen}}, \bibinfo {author}
  {\bibfnamefont {H.}~\bibnamefont {Kurebayashi}},\ and\ \bibinfo {author}
  {\bibfnamefont {W.~R.}\ \bibnamefont {Branford}},\ }\bibfield  {title}
  {\bibinfo {title} {Tunable magnetization dynamics in artificial spin ice via
  shape anisotropy modification},\ }\href@noop {} {\bibfield  {journal}
  {\bibinfo  {journal} {Phys. Rev. B}\ }\textbf {\bibinfo {volume} {100}},\
  \bibinfo {pages} {054433} (\bibinfo {year} {2019})}\BibitemShut {NoStop}%
\bibitem [{\citenamefont {Vanstone}\ \emph {et~al.}(2022)\citenamefont
  {Vanstone}, \citenamefont {Gartside}, \citenamefont {Stenning}, \citenamefont
  {Dion}, \citenamefont {Arroo},\ and\ \citenamefont
  {Branford}}]{Vanstone2022}%
  \BibitemOpen
  \bibfield  {author} {\bibinfo {author} {\bibfnamefont {A.}~\bibnamefont
  {Vanstone}}, \bibinfo {author} {\bibfnamefont {J.~C.}\ \bibnamefont
  {Gartside}}, \bibinfo {author} {\bibfnamefont {K.~D.}\ \bibnamefont
  {Stenning}}, \bibinfo {author} {\bibfnamefont {T.}~\bibnamefont {Dion}},
  \bibinfo {author} {\bibfnamefont {D.~M.}\ \bibnamefont {Arroo}},\ and\
  \bibinfo {author} {\bibfnamefont {W.~R.}\ \bibnamefont {Branford}},\
  }\bibfield  {title} {\bibinfo {title} {Spectral fingerprinting: microstate
  readout via remanence ferromagnetic resonance in artificial spin ice},\
  }\href {https://doi.org/10.1088/1367-2630/ac608b} {\bibfield  {journal}
  {\bibinfo  {journal} {New Journal of Physics}\ }\textbf {\bibinfo {volume}
  {24}},\ \bibinfo {pages} {043017} (\bibinfo {year} {2022})}\BibitemShut
  {NoStop}%
\bibitem [{\citenamefont {Lendinez}\ \emph {et~al.}(2023)\citenamefont
  {Lendinez}, \citenamefont {Kaffash}, \citenamefont {Heinonen}, \citenamefont
  {Gliga}, \citenamefont {Iacocca},\ and\ \citenamefont
  {Jungfleisch}}]{Lendinez2023}%
  \BibitemOpen
  \bibfield  {author} {\bibinfo {author} {\bibfnamefont {S.}~\bibnamefont
  {Lendinez}}, \bibinfo {author} {\bibfnamefont {M.~T.}\ \bibnamefont
  {Kaffash}}, \bibinfo {author} {\bibfnamefont {O.~G.}\ \bibnamefont
  {Heinonen}}, \bibinfo {author} {\bibfnamefont {S.}~\bibnamefont {Gliga}},
  \bibinfo {author} {\bibfnamefont {E.}~\bibnamefont {Iacocca}},\ and\ \bibinfo
  {author} {\bibfnamefont {M.~B.}\ \bibnamefont {Jungfleisch}},\ }\bibfield
  {title} {\bibinfo {title} {Nonlinear multi-magnon scattering in artificial
  spin ice},\ }\href {https://www.nature.com/articles/s41467-023-38992-7}
  {\bibfield  {journal} {\bibinfo  {journal} {Nature Communications}\ }\textbf
  {\bibinfo {volume} {14}},\ \bibinfo {pages} {3419} (\bibinfo {year}
  {2023})}\BibitemShut {NoStop}%
\bibitem [{\citenamefont {Alatteili}\ \emph {et~al.}(2024)\citenamefont
  {Alatteili}, \citenamefont {Roxburgh},\ and\ \citenamefont
  {Iacocca}}]{Alatteili2024}%
  \BibitemOpen
  \bibfield  {author} {\bibinfo {author} {\bibfnamefont {G.}~\bibnamefont
  {Alatteili}}, \bibinfo {author} {\bibfnamefont {A.}~\bibnamefont
  {Roxburgh}},\ and\ \bibinfo {author} {\bibfnamefont {E.}~\bibnamefont
  {Iacocca}},\ }\bibfield  {title} {\bibinfo {title} {Ferromagnetic resonance
  in three-dimensional tilted-square artificial spin ices},\ }\href
  {https://doi.org/10.1103/PhysRevB.110.144406} {\bibfield  {journal} {\bibinfo
   {journal} {Phys. Rev. B}\ }\textbf {\bibinfo {volume} {110}},\ \bibinfo
  {pages} {144406} (\bibinfo {year} {2024})}\BibitemShut {NoStop}%
\bibitem [{\citenamefont {Gartside}\ \emph {et~al.}(2022)\citenamefont
  {Gartside}, \citenamefont {Stenning}, \citenamefont {Vanstone}, \citenamefont
  {Holder}, \citenamefont {Arroo}, \citenamefont {Dion}, \citenamefont
  {Caravelli}, \citenamefont {Kurebayashi},\ and\ \citenamefont
  {Branford}}]{Gartside2022}%
  \BibitemOpen
  \bibfield  {author} {\bibinfo {author} {\bibfnamefont {J.~C.}\ \bibnamefont
  {Gartside}}, \bibinfo {author} {\bibfnamefont {K.~D.}\ \bibnamefont
  {Stenning}}, \bibinfo {author} {\bibfnamefont {A.}~\bibnamefont {Vanstone}},
  \bibinfo {author} {\bibfnamefont {H.~H.}\ \bibnamefont {Holder}}, \bibinfo
  {author} {\bibfnamefont {D.~M.}\ \bibnamefont {Arroo}}, \bibinfo {author}
  {\bibfnamefont {T.}~\bibnamefont {Dion}}, \bibinfo {author} {\bibfnamefont
  {F.}~\bibnamefont {Caravelli}}, \bibinfo {author} {\bibfnamefont
  {H.}~\bibnamefont {Kurebayashi}},\ and\ \bibinfo {author} {\bibfnamefont
  {W.~R.}\ \bibnamefont {Branford}},\ }\bibfield  {title} {\bibinfo {title}
  {Reconfigurable training and reservoir computing in an artificial spin-vortex
  ice via spin-wave fingerprinting},\ }\href
  {https://doi.org/10.1038/s41565-022-01091-7} {\bibfield  {journal} {\bibinfo
  {journal} {Nature Nanotechnology}\ }\textbf {\bibinfo {volume} {17}},\
  \bibinfo {pages} {460} (\bibinfo {year} {2022})}\BibitemShut {NoStop}%
\bibitem [{\citenamefont {Lee}\ \emph {et~al.}(2024)\citenamefont {Lee},
  \citenamefont {Wei}, \citenamefont {Stenning}, \citenamefont {Gartside},
  \citenamefont {Perstwood}, \citenamefont {Seki}, \citenamefont {Ageel},
  \citenamefont {Karube}, \citenamefont {Kanazawa}, \citenamefont {Taguchi},
  \citenamefont {Back}, \citenamefont {Tokura}, \citenamefont {Branford},\ and\
  \citenamefont {Kurebayashi}}]{Lee2024}%
  \BibitemOpen
  \bibfield  {author} {\bibinfo {author} {\bibfnamefont {O.}~\bibnamefont
  {Lee}}, \bibinfo {author} {\bibfnamefont {T.}~\bibnamefont {Wei}}, \bibinfo
  {author} {\bibfnamefont {K.~D.}\ \bibnamefont {Stenning}}, \bibinfo {author}
  {\bibfnamefont {J.~C.}\ \bibnamefont {Gartside}}, \bibinfo {author}
  {\bibfnamefont {D.}~\bibnamefont {Perstwood}}, \bibinfo {author}
  {\bibfnamefont {S.}~\bibnamefont {Seki}}, \bibinfo {author} {\bibfnamefont
  {A.}~\bibnamefont {Ageel}}, \bibinfo {author} {\bibfnamefont
  {K.}~\bibnamefont {Karube}}, \bibinfo {author} {\bibfnamefont
  {N.}~\bibnamefont {Kanazawa}}, \bibinfo {author} {\bibfnamefont
  {Y.}~\bibnamefont {Taguchi}}, \bibinfo {author} {\bibfnamefont
  {C.}~\bibnamefont {Back}}, \bibinfo {author} {\bibfnamefont {Y.}~\bibnamefont
  {Tokura}}, \bibinfo {author} {\bibfnamefont {W.~R.}\ \bibnamefont
  {Branford}},\ and\ \bibinfo {author} {\bibfnamefont {H.}~\bibnamefont
  {Kurebayashi}},\ }\bibfield  {title} {\bibinfo {title} {Task-adaptive
  physical reservoir computing},\ }\href
  {https://doi.org/10.1038/s41563-023-01698-8} {\bibfield  {journal} {\bibinfo
  {journal} {Nature Materials}\ }\textbf {\bibinfo {volume} {23}},\ \bibinfo
  {pages} {79} (\bibinfo {year} {2024})}\BibitemShut {NoStop}%
\bibitem [{\citenamefont {Stenning}\ \emph {et~al.}(2024)\citenamefont
  {Stenning}, \citenamefont {Gartside}, \citenamefont {Manneschi},
  \citenamefont {Cheung}, \citenamefont {Chen}, \citenamefont {Vanstone},
  \citenamefont {Love}, \citenamefont {Holder}, \citenamefont {Caravelli},
  \citenamefont {Kurebayashi}, \citenamefont {Everschor-Sitte}, \citenamefont
  {Vasilaki},\ and\ \citenamefont {Branford}}]{Stenning2024}%
  \BibitemOpen
  \bibfield  {author} {\bibinfo {author} {\bibfnamefont {K.~D.}\ \bibnamefont
  {Stenning}}, \bibinfo {author} {\bibfnamefont {J.~C.}\ \bibnamefont
  {Gartside}}, \bibinfo {author} {\bibfnamefont {L.}~\bibnamefont {Manneschi}},
  \bibinfo {author} {\bibfnamefont {C.~T.~S.}\ \bibnamefont {Cheung}}, \bibinfo
  {author} {\bibfnamefont {T.}~\bibnamefont {Chen}}, \bibinfo {author}
  {\bibfnamefont {A.}~\bibnamefont {Vanstone}}, \bibinfo {author}
  {\bibfnamefont {J.}~\bibnamefont {Love}}, \bibinfo {author} {\bibfnamefont
  {H.}~\bibnamefont {Holder}}, \bibinfo {author} {\bibfnamefont
  {F.}~\bibnamefont {Caravelli}}, \bibinfo {author} {\bibfnamefont
  {H.}~\bibnamefont {Kurebayashi}}, \bibinfo {author} {\bibfnamefont
  {K.}~\bibnamefont {Everschor-Sitte}}, \bibinfo {author} {\bibfnamefont
  {E.}~\bibnamefont {Vasilaki}},\ and\ \bibinfo {author} {\bibfnamefont
  {W.~R.}\ \bibnamefont {Branford}},\ }\bibfield  {title} {\bibinfo {title}
  {Neuromorphic overparameterisation and few-shot learning in multilayer
  physical neural networks},\ }\href
  {https://doi.org/10.1038/s41467-024-50633-1} {\bibfield  {journal} {\bibinfo
  {journal} {Nature Communications}\ }\textbf {\bibinfo {volume} {15}},\
  \bibinfo {pages} {7377} (\bibinfo {year} {2024})}\BibitemShut {NoStop}%
\bibitem [{\citenamefont {Alatteili}\ \emph {et~al.}(2023)\citenamefont
  {Alatteili}, \citenamefont {Martinez}, \citenamefont {Roxburgh},
  \citenamefont {Gartside}, \citenamefont {Heinonen}, \citenamefont {Gliga},\
  and\ \citenamefont {Iacocca}}]{Alatteili2023}%
  \BibitemOpen
  \bibfield  {author} {\bibinfo {author} {\bibfnamefont {G.}~\bibnamefont
  {Alatteili}}, \bibinfo {author} {\bibfnamefont {V.}~\bibnamefont {Martinez}},
  \bibinfo {author} {\bibfnamefont {A.}~\bibnamefont {Roxburgh}}, \bibinfo
  {author} {\bibfnamefont {J.}~\bibnamefont {Gartside}}, \bibinfo {author}
  {\bibfnamefont {O.}~\bibnamefont {Heinonen}}, \bibinfo {author}
  {\bibfnamefont {S.}~\bibnamefont {Gliga}},\ and\ \bibinfo {author}
  {\bibfnamefont {E.}~\bibnamefont {Iacocca}},\ }\bibfield  {title} {\bibinfo
  {title} {G\ae{}nice: A general model for magnon band structure of artificial
  spin ices},\ }\href {https://doi.org/10.1016/j.jmmm.2023.171603} {\bibfield
  {journal} {\bibinfo  {journal} {Journal of Magnetism and Magnetic Materials}\
  }\textbf {\bibinfo {volume} {589}},\ \bibinfo {pages} {171603} (\bibinfo
  {year} {2023})}\BibitemShut {NoStop}%
\bibitem [{\citenamefont {Alatteili}\ \emph {et~al.}(2025)\citenamefont
  {Alatteili}, \citenamefont {Scafuri},\ and\ \citenamefont
  {Iacocca}}]{Alatteili2025}%
  \BibitemOpen
  \bibfield  {author} {\bibinfo {author} {\bibfnamefont {G.}~\bibnamefont
  {Alatteili}}, \bibinfo {author} {\bibfnamefont {L.}~\bibnamefont {Scafuri}},\
  and\ \bibinfo {author} {\bibfnamefont {E.}~\bibnamefont {Iacocca}},\
  }\bibfield  {title} {\bibinfo {title} {Symmetry-broken magneto-toroidal
  artificial spin ice: magnetization states and dynamics},\ }\href@noop {}
  {\bibfield  {journal} {\bibinfo  {journal} {arXiv:2505.12011}\ } (\bibinfo
  {year} {2025})}\BibitemShut {NoStop}%
\bibitem [{\citenamefont {Scafuri}\ \emph {et~al.}(2025)\citenamefont
  {Scafuri}, \citenamefont {Bozhko},\ and\ \citenamefont
  {Iacocca}}]{Scafuri2025}%
  \BibitemOpen
  \bibfield  {author} {\bibinfo {author} {\bibfnamefont {L.~A.}\ \bibnamefont
  {Scafuri}}, \bibinfo {author} {\bibfnamefont {D.~A.}\ \bibnamefont
  {Bozhko}},\ and\ \bibinfo {author} {\bibfnamefont {E.}~\bibnamefont
  {Iacocca}},\ }\bibfield  {title} {\bibinfo {title} {Wave dynamics in a
  macroscopic square artificial spin ice},\ }\href
  {https://doi.org/10.1103/PhysRevApplied.23.054093} {\bibfield  {journal}
  {\bibinfo  {journal} {Phys. Rev. Appl.}\ }\textbf {\bibinfo {volume} {23}},\
  \bibinfo {pages} {054093} (\bibinfo {year} {2025})}\BibitemShut {NoStop}%
\bibitem [{\citenamefont {Peroor}\ \emph {et~al.}(2025)\citenamefont {Peroor},
  \citenamefont {Scafuri}, \citenamefont {Bozhko},\ and\ \citenamefont
  {Iacocca}}]{Peroor2025}%
  \BibitemOpen
  \bibfield  {author} {\bibinfo {author} {\bibfnamefont {R.~R.}\ \bibnamefont
  {Peroor}}, \bibinfo {author} {\bibfnamefont {L.~A.}\ \bibnamefont {Scafuri}},
  \bibinfo {author} {\bibfnamefont {D.~A.}\ \bibnamefont {Bozhko}},\ and\
  \bibinfo {author} {\bibfnamefont {E.}~\bibnamefont {Iacocca}},\ }\bibfield
  {title} {\bibinfo {title} {Frequency comb in a macroscopic mechanomagnetic
  artificial spin ice},\ }\href
  {https://doi.org/10.1103/PhysRevApplied.23.044010} {\bibfield  {journal}
  {\bibinfo  {journal} {Phys. Rev. Appl.}\ }\textbf {\bibinfo {volume} {23}},\
  \bibinfo {pages} {044010} (\bibinfo {year} {2025})}\BibitemShut {NoStop}%
\bibitem [{\citenamefont {\"{O}stman}\ \emph {et~al.}(2018)\citenamefont
  {\"{O}stman}, \citenamefont {Stopfel}, \citenamefont {Chioar}, \citenamefont
  {Arnalds}, \citenamefont {Stein}, \citenamefont {Kapaklis},\ and\
  \citenamefont {Hj\"{o}rvarsson}}]{Ostman2018}%
  \BibitemOpen
  \bibfield  {author} {\bibinfo {author} {\bibfnamefont {E.}~\bibnamefont
  {\"{O}stman}}, \bibinfo {author} {\bibfnamefont {H.}~\bibnamefont {Stopfel}},
  \bibinfo {author} {\bibfnamefont {I.-A.}\ \bibnamefont {Chioar}}, \bibinfo
  {author} {\bibfnamefont {U.~B.}\ \bibnamefont {Arnalds}}, \bibinfo {author}
  {\bibfnamefont {A.}~\bibnamefont {Stein}}, \bibinfo {author} {\bibfnamefont
  {V.}~\bibnamefont {Kapaklis}},\ and\ \bibinfo {author} {\bibfnamefont
  {B.}~\bibnamefont {Hj\"{o}rvarsson}},\ }\bibfield  {title} {\bibinfo {title}
  {Interaction modifiers in artificial spin ices},\ }\href
  {https://doi.org/10.1038/s41567-017-0027-2} {\bibfield  {journal} {\bibinfo
  {journal} {Nature Physics}\ }\textbf {\bibinfo {volume} {14}},\ \bibinfo
  {pages} {375–379} (\bibinfo {year} {2018})}\BibitemShut {NoStop}%
\bibitem [{\citenamefont {Berchialla}\ \emph {et~al.}(2024)\citenamefont
  {Berchialla}, \citenamefont {Macauley},\ and\ \citenamefont
  {Heyderman}}]{Berchialla2024}%
  \BibitemOpen
  \bibfield  {author} {\bibinfo {author} {\bibfnamefont {L.}~\bibnamefont
  {Berchialla}}, \bibinfo {author} {\bibfnamefont {G.~M.}\ \bibnamefont
  {Macauley}},\ and\ \bibinfo {author} {\bibfnamefont {L.~J.}\ \bibnamefont
  {Heyderman}},\ }\bibfield  {title} {\bibinfo {title} {Focus on
  three-dimensional artificial spin ice},\ }\bibfield  {journal} {\bibinfo
  {journal} {Applied Physics Letters}\ }\textbf {\bibinfo {volume} {125}},\
  \href {https://doi.org/10.1063/5.0229120} {10.1063/5.0229120} (\bibinfo
  {year} {2024})\BibitemShut {NoStop}%
\bibitem [{\citenamefont {Sultana}\ \emph {et~al.}(2025)\citenamefont
  {Sultana}, \citenamefont {Mondal}, \citenamefont {Bhat}, \citenamefont
  {Stenning}, \citenamefont {Li}, \citenamefont {Arroo}, \citenamefont
  {Vasdev}, \citenamefont {McCarter}, \citenamefont {De~Long}, \citenamefont
  {Hastings}, \citenamefont {Gartside},\ and\ \citenamefont
  {Jungfleisch}}]{Sultana2025}%
  \BibitemOpen
  \bibfield  {author} {\bibinfo {author} {\bibfnamefont {R.}~\bibnamefont
  {Sultana}}, \bibinfo {author} {\bibfnamefont {A.~K.}\ \bibnamefont {Mondal}},
  \bibinfo {author} {\bibfnamefont {V.~S.}\ \bibnamefont {Bhat}}, \bibinfo
  {author} {\bibfnamefont {K.}~\bibnamefont {Stenning}}, \bibinfo {author}
  {\bibfnamefont {Y.}~\bibnamefont {Li}}, \bibinfo {author} {\bibfnamefont
  {D.~M.}\ \bibnamefont {Arroo}}, \bibinfo {author} {\bibfnamefont
  {A.}~\bibnamefont {Vasdev}}, \bibinfo {author} {\bibfnamefont {M.~R.}\
  \bibnamefont {McCarter}}, \bibinfo {author} {\bibfnamefont {L.~E.}\
  \bibnamefont {De~Long}}, \bibinfo {author} {\bibfnamefont {J.~T.}\
  \bibnamefont {Hastings}}, \bibinfo {author} {\bibfnamefont {J.~C.}\
  \bibnamefont {Gartside}},\ and\ \bibinfo {author} {\bibfnamefont {M.~B.}\
  \bibnamefont {Jungfleisch}},\ }\bibfield  {title} {\bibinfo {title} {Ice
  sculpting: An artificial spin ice tutorial on controlling microstate and
  geometry for magnonics and neuromorphic computing},\ }\href
  {https://doi.org/10.1063/5.0274799} {\bibfield  {journal} {\bibinfo
  {journal} {Journal of Applied Physics}\ }\textbf {\bibinfo {volume} {138}},\
  \bibinfo {pages} {061101} (\bibinfo {year} {2025})},\ \Eprint
  {https://arxiv.org/abs/https://pubs.aip.org/aip/jap/article-pdf/doi/10.1063/5.0274799/20636646/061101\_1\_5.0274799.pdf}
  {https://pubs.aip.org/aip/jap/article-pdf/doi/10.1063/5.0274799/20636646/061101\_1\_5.0274799.pdf}
  \BibitemShut {NoStop}%
\bibitem [{\citenamefont {Micaletti}\ \emph
  {et~al.}(2025{\natexlab{a}})\citenamefont {Micaletti}, \citenamefont
  {Roxburgh}, \citenamefont {Iacocca}, \citenamefont {Marzolla},\ and\
  \citenamefont {Montoncello}}]{Micaletti2025}%
  \BibitemOpen
  \bibfield  {author} {\bibinfo {author} {\bibfnamefont {P.}~\bibnamefont
  {Micaletti}}, \bibinfo {author} {\bibfnamefont {A.}~\bibnamefont {Roxburgh}},
  \bibinfo {author} {\bibfnamefont {E.}~\bibnamefont {Iacocca}}, \bibinfo
  {author} {\bibfnamefont {M.}~\bibnamefont {Marzolla}},\ and\ \bibinfo
  {author} {\bibfnamefont {F.}~\bibnamefont {Montoncello}},\ }\bibfield
  {title} {\bibinfo {title} {A sinusoidal magnetization distribution as an
  original way to generate a versatile magnonic crystal for magnon
  propagation},\ }\href
  {https://doi.org/https://doi.org/10.1016/j.jmmm.2025.172959} {\bibfield
  {journal} {\bibinfo  {journal} {Journal of Magnetism and Magnetic Materials}\
  }\textbf {\bibinfo {volume} {622}},\ \bibinfo {pages} {172959} (\bibinfo
  {year} {2025}{\natexlab{a}})}\BibitemShut {NoStop}%
\bibitem [{\citenamefont {Micaletti}\ \emph
  {et~al.}(2025{\natexlab{b}})\citenamefont {Micaletti}, \citenamefont
  {Roxburgh}, \citenamefont {Iacocca}, \citenamefont {Marzolla},\ and\
  \citenamefont {Montoncello}}]{Micaletti2025b}%
  \BibitemOpen
  \bibfield  {author} {\bibinfo {author} {\bibfnamefont {P.}~\bibnamefont
  {Micaletti}}, \bibinfo {author} {\bibfnamefont {A.}~\bibnamefont {Roxburgh}},
  \bibinfo {author} {\bibfnamefont {E.}~\bibnamefont {Iacocca}}, \bibinfo
  {author} {\bibfnamefont {M.}~\bibnamefont {Marzolla}},\ and\ \bibinfo
  {author} {\bibfnamefont {F.}~\bibnamefont {Montoncello}},\ }\bibfield
  {title} {\bibinfo {title} {Magnonic analog of a metal-to-insulator transition
  in a multiferroic heterostructure},\ }\href
  {https://doi.org/10.1063/5.0250690} {\bibfield  {journal} {\bibinfo
  {journal} {Journal of Applied Physics}\ }\textbf {\bibinfo {volume} {137}},\
  \bibinfo {pages} {153906} (\bibinfo {year} {2025}{\natexlab{b}})}\BibitemShut
  {NoStop}%
\bibitem [{\citenamefont {Dion}\ \emph {et~al.}(2024)\citenamefont {Dion},
  \citenamefont {Stenning}, \citenamefont {Vanstone}, \citenamefont {Holder},
  \citenamefont {Sultana}, \citenamefont {Alatteili}, \citenamefont {Martinez},
  \citenamefont {Kaffash}, \citenamefont {T.}, \citenamefont {Kurebayashi},
  \citenamefont {Branford}, \citenamefont {Iacocca}, \citenamefont
  {Jungfleisch},\ and\ \citenamefont {Gartside}}]{Dion2024}%
  \BibitemOpen
  \bibfield  {author} {\bibinfo {author} {\bibfnamefont {T.}~\bibnamefont
  {Dion}}, \bibinfo {author} {\bibfnamefont {K.~D.}\ \bibnamefont {Stenning}},
  \bibinfo {author} {\bibfnamefont {A.}~\bibnamefont {Vanstone}}, \bibinfo
  {author} {\bibfnamefont {H.~H.}\ \bibnamefont {Holder}}, \bibinfo {author}
  {\bibfnamefont {R.}~\bibnamefont {Sultana}}, \bibinfo {author} {\bibfnamefont
  {G.}~\bibnamefont {Alatteili}}, \bibinfo {author} {\bibfnamefont
  {V.}~\bibnamefont {Martinez}}, \bibinfo {author} {\bibfnamefont {M.~T.}\
  \bibnamefont {Kaffash}}, \bibinfo {author} {\bibfnamefont {K.}~\bibnamefont
  {T.}}, \bibinfo {author} {\bibfnamefont {H.}~\bibnamefont {Kurebayashi}},
  \bibinfo {author} {\bibfnamefont {W.~R.}\ \bibnamefont {Branford}}, \bibinfo
  {author} {\bibfnamefont {E.}~\bibnamefont {Iacocca}}, \bibinfo {author}
  {\bibfnamefont {B.~M.}\ \bibnamefont {Jungfleisch}},\ and\ \bibinfo {author}
  {\bibfnamefont {J.~C.}\ \bibnamefont {Gartside}},\ }\bibfield  {title}
  {\bibinfo {title} {Ultrastrong magnon-magnon coupling and chiral spin texture
  control in a dipolar 3d multilayered artificial spin-vortex ice},\ }\href
  {https://doi.org/10.1038/s41467-024-48080-z} {\bibfield  {journal} {\bibinfo
  {journal} {Nature Communications}\ }\textbf {\bibinfo {volume} {15}},\
  \bibinfo {pages} {4077} (\bibinfo {year} {2024})}\BibitemShut {NoStop}%
\bibitem [{\citenamefont {Nembach}\ \emph {et~al.}(2015)\citenamefont
  {Nembach}, \citenamefont {Shaw}, \citenamefont {Weller}, \citenamefont
  {Ju\'{e}},\ and\ \citenamefont {Silva}}]{Nembach2015}%
  \BibitemOpen
  \bibfield  {author} {\bibinfo {author} {\bibfnamefont {H.~T.}\ \bibnamefont
  {Nembach}}, \bibinfo {author} {\bibfnamefont {J.~M.}\ \bibnamefont {Shaw}},
  \bibinfo {author} {\bibfnamefont {M.}~\bibnamefont {Weller}}, \bibinfo
  {author} {\bibfnamefont {E.}~\bibnamefont {Ju\'{e}}},\ and\ \bibinfo {author}
  {\bibfnamefont {T.~J.}\ \bibnamefont {Silva}},\ }\bibfield  {title} {\bibinfo
  {title} {Linear relation between heisenberg exchange and interfacial
  dzyaloshinskii-moriya interaction in metal films},\ }\href@noop {} {\bibfield
   {journal} {\bibinfo  {journal} {Nature Physics}\ }\textbf {\bibinfo {volume}
  {11}},\ \bibinfo {pages} {825} (\bibinfo {year} {2015})}\BibitemShut
  {NoStop}%
\bibitem [{\citenamefont {Kim}\ \emph {et~al.}(2016)\citenamefont {Kim},
  \citenamefont {Stamps},\ and\ \citenamefont {Camley}}]{Kim2016c}%
  \BibitemOpen
  \bibfield  {author} {\bibinfo {author} {\bibfnamefont {J.-V.}\ \bibnamefont
  {Kim}}, \bibinfo {author} {\bibfnamefont {R.~L.}\ \bibnamefont {Stamps}},\
  and\ \bibinfo {author} {\bibfnamefont {R.~E.}\ \bibnamefont {Camley}},\
  }\bibfield  {title} {\bibinfo {title} {Spin wave power flow and caustics in
  ultrathin ferromagnets with the dzyaloshinskii-moriya interaction},\ }\href
  {https://doi.org/10.1103/PhysRevLett.117.197204} {\bibfield  {journal}
  {\bibinfo  {journal} {Phys. Rev. Lett.}\ }\textbf {\bibinfo {volume} {117}},\
  \bibinfo {pages} {197204} (\bibinfo {year} {2016})}\BibitemShut {NoStop}%
\bibitem [{\citenamefont {Gallardo}\ \emph {et~al.}(2019)\citenamefont
  {Gallardo}, \citenamefont {Cort\'es-Ortu\~no}, \citenamefont {Schneider},
  \citenamefont {Rold\'an-Molina}, \citenamefont {Ma}, \citenamefont
  {Troncoso}, \citenamefont {Lenz}, \citenamefont {Fangohr}, \citenamefont
  {Lindner},\ and\ \citenamefont {Landeros}}]{Gallardo2019}%
  \BibitemOpen
  \bibfield  {author} {\bibinfo {author} {\bibfnamefont {R.~A.}\ \bibnamefont
  {Gallardo}}, \bibinfo {author} {\bibfnamefont {D.}~\bibnamefont
  {Cort\'es-Ortu\~no}}, \bibinfo {author} {\bibfnamefont {T.}~\bibnamefont
  {Schneider}}, \bibinfo {author} {\bibfnamefont {A.}~\bibnamefont
  {Rold\'an-Molina}}, \bibinfo {author} {\bibfnamefont {F.}~\bibnamefont {Ma}},
  \bibinfo {author} {\bibfnamefont {R.~E.}\ \bibnamefont {Troncoso}}, \bibinfo
  {author} {\bibfnamefont {K.}~\bibnamefont {Lenz}}, \bibinfo {author}
  {\bibfnamefont {H.}~\bibnamefont {Fangohr}}, \bibinfo {author} {\bibfnamefont
  {J.}~\bibnamefont {Lindner}},\ and\ \bibinfo {author} {\bibfnamefont
  {P.}~\bibnamefont {Landeros}},\ }\bibfield  {title} {\bibinfo {title} {Flat
  bands, indirect gaps, and unconventional spin-wave behavior induced by a
  periodic dzyaloshinskii-moriya interaction},\ }\href
  {https://doi.org/10.1103/PhysRevLett.122.067204} {\bibfield  {journal}
  {\bibinfo  {journal} {Phys. Rev. Lett.}\ }\textbf {\bibinfo {volume} {122}},\
  \bibinfo {pages} {067204} (\bibinfo {year} {2019})}\BibitemShut {NoStop}%
\bibitem [{\citenamefont {Zingsem}\ \emph {et~al.}(2019)\citenamefont
  {Zingsem}, \citenamefont {Farle}, \citenamefont {Stamps},\ and\ \citenamefont
  {Camley}}]{Zingsem2019}%
  \BibitemOpen
  \bibfield  {author} {\bibinfo {author} {\bibfnamefont {B.~W.}\ \bibnamefont
  {Zingsem}}, \bibinfo {author} {\bibfnamefont {M.}~\bibnamefont {Farle}},
  \bibinfo {author} {\bibfnamefont {R.~L.}\ \bibnamefont {Stamps}},\ and\
  \bibinfo {author} {\bibfnamefont {R.~E.}\ \bibnamefont {Camley}},\ }\bibfield
   {title} {\bibinfo {title} {Unusual nature of confined modes in a chiral
  system: Directional transport in standing waves},\ }\href
  {https://doi.org/10.1103/PhysRevB.99.214429} {\bibfield  {journal} {\bibinfo
  {journal} {Phys. Rev. B}\ }\textbf {\bibinfo {volume} {99}},\ \bibinfo
  {pages} {214429} (\bibinfo {year} {2019})}\BibitemShut {NoStop}%
\bibitem [{\citenamefont {Schoen}\ \emph {et~al.}(2016)\citenamefont {Schoen},
  \citenamefont {Thonig}, \citenamefont {Schneider}, \citenamefont {Silva},
  \citenamefont {Nembach}, \citenamefont {Eriksson}, \citenamefont {Karis},\
  and\ \citenamefont {Shaw}}]{Schoen2016}%
  \BibitemOpen
  \bibfield  {author} {\bibinfo {author} {\bibfnamefont {M.~A.}\ \bibnamefont
  {Schoen}}, \bibinfo {author} {\bibfnamefont {D.}~\bibnamefont {Thonig}},
  \bibinfo {author} {\bibfnamefont {M.~L.}\ \bibnamefont {Schneider}}, \bibinfo
  {author} {\bibfnamefont {T.}~\bibnamefont {Silva}}, \bibinfo {author}
  {\bibfnamefont {H.~T.}\ \bibnamefont {Nembach}}, \bibinfo {author}
  {\bibfnamefont {O.}~\bibnamefont {Eriksson}}, \bibinfo {author}
  {\bibfnamefont {O.}~\bibnamefont {Karis}},\ and\ \bibinfo {author}
  {\bibfnamefont {J.~M.}\ \bibnamefont {Shaw}},\ }\bibfield  {title} {\bibinfo
  {title} {Ultra-low magnetic damping of a metallic ferromagnet},\ }\href@noop
  {} {\bibfield  {journal} {\bibinfo  {journal} {Nature Physics}\ }\textbf
  {\bibinfo {volume} {12}},\ \bibinfo {pages} {839} (\bibinfo {year}
  {2016})}\BibitemShut {NoStop}%
\bibitem [{\citenamefont {Vansteenkiste}\ \emph {et~al.}(2014)\citenamefont
  {Vansteenkiste}, \citenamefont {Leliaert}, \citenamefont {Dvornik},
  \citenamefont {Helsen}, \citenamefont {Garcia-Sanchez},\ and\ \citenamefont
  {Van~Waeyenberge}}]{Vansteenkiste2014}%
  \BibitemOpen
  \bibfield  {author} {\bibinfo {author} {\bibfnamefont {A.}~\bibnamefont
  {Vansteenkiste}}, \bibinfo {author} {\bibfnamefont {J.}~\bibnamefont
  {Leliaert}}, \bibinfo {author} {\bibfnamefont {M.}~\bibnamefont {Dvornik}},
  \bibinfo {author} {\bibfnamefont {M.}~\bibnamefont {Helsen}}, \bibinfo
  {author} {\bibfnamefont {F.}~\bibnamefont {Garcia-Sanchez}},\ and\ \bibinfo
  {author} {\bibfnamefont {B.}~\bibnamefont {Van~Waeyenberge}},\ }\bibfield
  {title} {\bibinfo {title} {The design and verification of mumax3},\
  }\href@noop {} {\bibfield  {journal} {\bibinfo  {journal} {AIP Advances}\
  }\textbf {\bibinfo {volume} {4}},\ \bibinfo {pages} {107133} (\bibinfo {year}
  {2014})}\BibitemShut {NoStop}%
\bibitem [{\citenamefont {Strandqvist}\ \emph {et~al.}(2025)\citenamefont
  {Strandqvist}, \citenamefont {Fitez}, \citenamefont {Heinonen},\ and\
  \citenamefont {Schiffer}}]{Strandqvist2025}%
  \BibitemOpen
  \bibfield  {author} {\bibinfo {author} {\bibfnamefont {N.}~\bibnamefont
  {Strandqvist}}, \bibinfo {author} {\bibfnamefont {G.}~\bibnamefont {Fitez}},
  \bibinfo {author} {\bibfnamefont {O.}~\bibnamefont {Heinonen}},\ and\
  \bibinfo {author} {\bibfnamefont {P.}~\bibnamefont {Schiffer}},\ }\bibfield
  {title} {\bibinfo {title} {Nanomagnet shape effects on magnetic reversal in
  artificial spin ice},\ }\href {https://doi.org/10.1103/PhysRevB.111.184420}
  {\bibfield  {journal} {\bibinfo  {journal} {Phys. Rev. B}\ }\textbf {\bibinfo
  {volume} {111}},\ \bibinfo {pages} {184420} (\bibinfo {year}
  {2025})}\BibitemShut {NoStop}%
\bibitem [{\citenamefont {Vanga}\ and\ \citenamefont
  {Iacocca}(2025)}]{Vanga2025}%
  \BibitemOpen
  \bibfield  {author} {\bibinfo {author} {\bibfnamefont {V.}~\bibnamefont
  {Vanga}}\ and\ \bibinfo {author} {\bibfnamefont {E.}~\bibnamefont
  {Iacocca}},\ }\bibfield  {title} {\bibinfo {title} {Dependence of the
  ferromagnetic resonance of artificial spin ices on nanomagnet shape},\ }\href
  {https://doi.org/10.1103/PhysRevB.111.184421} {\bibfield  {journal} {\bibinfo
   {journal} {Phys. Rev. B}\ }\textbf {\bibinfo {volume} {111}},\ \bibinfo
  {pages} {184421} (\bibinfo {year} {2025})}\BibitemShut {NoStop}%
\bibitem [{\citenamefont {Iacocca}\ \emph {et~al.}(2020)\citenamefont
  {Iacocca}, \citenamefont {Gliga},\ and\ \citenamefont
  {Heinonen}}]{Iacocca2020}%
  \BibitemOpen
  \bibfield  {author} {\bibinfo {author} {\bibfnamefont {E.}~\bibnamefont
  {Iacocca}}, \bibinfo {author} {\bibfnamefont {S.}~\bibnamefont {Gliga}},\
  and\ \bibinfo {author} {\bibfnamefont {O.~G.}\ \bibnamefont {Heinonen}},\
  }\bibfield  {title} {\bibinfo {title} {Tailoring spin-wave channels in a
  reconfigurable artificial spin ice},\ }\href
  {https://doi.org/10.1103/PhysRevApplied.13.044047} {\bibfield  {journal}
  {\bibinfo  {journal} {Phys. Rev. Applied}\ }\textbf {\bibinfo {volume}
  {13}},\ \bibinfo {pages} {044047} (\bibinfo {year} {2020})}\BibitemShut
  {NoStop}%
\bibitem [{\citenamefont {Lendinez}\ \emph {et~al.}(2021)\citenamefont
  {Lendinez}, \citenamefont {Kaffash},\ and\ \citenamefont
  {Jungfleisch}}]{Lendinez2021}%
  \BibitemOpen
  \bibfield  {author} {\bibinfo {author} {\bibfnamefont {S.}~\bibnamefont
  {Lendinez}}, \bibinfo {author} {\bibfnamefont {M.~T.}\ \bibnamefont
  {Kaffash}},\ and\ \bibinfo {author} {\bibfnamefont {M.~B.}\ \bibnamefont
  {Jungfleisch}},\ }\bibfield  {title} {\bibinfo {title} {Emergent spin
  dynamics enabled by lattice interactions in a bicomponent artificial spin
  ice},\ }\href {https://pubs.acs.org/doi/10.1021/acs.nanolett.0c03729}
  {\bibfield  {journal} {\bibinfo  {journal} {Nano Lett.}\ }\textbf {\bibinfo
  {volume} {21}},\ \bibinfo {pages} {1921} (\bibinfo {year}
  {2021})}\BibitemShut {NoStop}%
\bibitem [{\citenamefont {Luo}\ \emph {et~al.}(2019)\citenamefont {Luo},
  \citenamefont {Dao}, \citenamefont {Hrabec}, \citenamefont {Vijayakumar},
  \citenamefont {Kleibert}, \citenamefont {Baumgartner}, \citenamefont {Kirk},
  \citenamefont {Cui}, \citenamefont {Savchenko}, \citenamefont {Krishnaswamy},
  \citenamefont {Heyderman},\ and\ \citenamefont {Gambardella}}]{Luo2019}%
  \BibitemOpen
  \bibfield  {author} {\bibinfo {author} {\bibfnamefont {Z.}~\bibnamefont
  {Luo}}, \bibinfo {author} {\bibfnamefont {T.~P.}\ \bibnamefont {Dao}},
  \bibinfo {author} {\bibfnamefont {A.}~\bibnamefont {Hrabec}}, \bibinfo
  {author} {\bibfnamefont {J.}~\bibnamefont {Vijayakumar}}, \bibinfo {author}
  {\bibfnamefont {A.}~\bibnamefont {Kleibert}}, \bibinfo {author}
  {\bibfnamefont {M.}~\bibnamefont {Baumgartner}}, \bibinfo {author}
  {\bibfnamefont {E.}~\bibnamefont {Kirk}}, \bibinfo {author} {\bibfnamefont
  {J.}~\bibnamefont {Cui}}, \bibinfo {author} {\bibfnamefont {T.}~\bibnamefont
  {Savchenko}}, \bibinfo {author} {\bibfnamefont {G.}~\bibnamefont
  {Krishnaswamy}}, \bibinfo {author} {\bibfnamefont {L.~J.}\ \bibnamefont
  {Heyderman}},\ and\ \bibinfo {author} {\bibfnamefont {P.}~\bibnamefont
  {Gambardella}},\ }\bibfield  {title} {\bibinfo {title} {Chirally coupled
  nanomagnets},\ }\href@noop {} {\bibfield  {journal} {\bibinfo  {journal}
  {Science}\ }\textbf {\bibinfo {volume} {363}},\ \bibinfo {pages} {1435}
  (\bibinfo {year} {2019})}\BibitemShut {NoStop}%
\bibitem [{\citenamefont {Iacocca}\ and\ \citenamefont
  {Heinonen}(2017)}]{Iacocca2017c}%
  \BibitemOpen
  \bibfield  {author} {\bibinfo {author} {\bibfnamefont {E.}~\bibnamefont
  {Iacocca}}\ and\ \bibinfo {author} {\bibfnamefont {O.}~\bibnamefont
  {Heinonen}},\ }\bibfield  {title} {\bibinfo {title} {Topologically nontrivial
  magnon bands in artificial square spin ices with dzyaloshinskii-moriya
  interaction},\ }\href@noop {} {\bibfield  {journal} {\bibinfo  {journal}
  {Phys. Rev. Applied}\ }\textbf {\bibinfo {volume} {8}},\ \bibinfo {pages}
  {034015} (\bibinfo {year} {2017})}\BibitemShut {NoStop}%
\bibitem [{OSF()}]{OSF}%
  \BibitemOpen
  \href@noop {} {}\bibinfo {note} {The data are available at
  https://osf.io/BUQHP/files/osfstorage under the folder ``Ferromagnetic
  resonance modes in trilayer artificial spin ices subject to interfacial
  Dzyaloshinskii-Moriya interaction''.}\BibitemShut {Stop}%
\end{thebibliography}
\end{document}